\PassOptionsToPackage{table,xcdraw}{xcolor}
\documentclass[11pt,a4paper]{article}
\pdfoutput=1

\makeatletter
\def\@fpheader{}
\makeatother

\usepackage{jheppub}
\usepackage{gensymb}
\DeclareSymbolFont{matha}{OML}{txmi}{m}{it}
\DeclareMathSymbol{\varv}{\mathord}{matha}{118}
\usepackage{amsmath}
\usepackage{amssymb}
\usepackage{mathrsfs}
\usepackage{graphicx}
\usepackage{subfigure}
\usepackage{pstricks}
\usepackage{bm}
\usepackage{pbox}
\usepackage{placeins}
\usepackage{booktabs}
\usepackage[T1]{fontenc}
\usepackage{footnote}
\usepackage{pdfpages}
\usepackage{hhline}
\usepackage{multirow}
\usepackage{multicol}
\usepackage{enumitem}
\usepackage[toc,page]{appendix}
\allowdisplaybreaks
\usepackage{comment}
\usepackage{float}
\usepackage{slashed}
\usepackage{xcolor}
\usepackage{textcomp}
\usepackage{multirow}
\usepackage[numbers]{natbib}
\usepackage{notoccite}
\usepackage{indentfirst}
\usepackage{jheppub}
\usepackage{gensymb}
\DeclareSymbolFont{matha}{OML}{txmi}{m}{it}
\DeclareMathSymbol{\varv}{\mathord}{matha}{118}
\usepackage{amsmath}
\usepackage{array}
\usepackage{amssymb}
\usepackage{graphicx}
\usepackage{subfigure}
\usepackage{pstricks}
\usepackage{bm}
\usepackage{pbox}
\usepackage{placeins}
\usepackage{booktabs}
\usepackage[T1]{fontenc}
\usepackage{footnote}
\usepackage{pdfpages}
\usepackage{hhline}
\usepackage{multirow}
\usepackage{multicol}
\usepackage{enumitem}
\usepackage[toc,page]{appendix}
\allowdisplaybreaks
\usepackage{comment}
\usepackage{float}
\usepackage{slashed}
\usepackage{xcolor}
\usepackage{textcomp}
\usepackage{multirow}
\usepackage[numbers]{natbib}
\usepackage{notoccite}
\usepackage{soul}
\usepackage{extarrows}
\usepackage{makecell}
\usepackage{longtable}
\usepackage{physics}
\usepackage{comment}
\usepackage{blindtext}
\usepackage{csquotes}
\usepackage{soul}
\usepackage{indentfirst}
\DeclareUnicodeCharacter{0304}{ }
\usepackage{rotating}
\usepackage{tabularx}
\usepackage{extarrows}
\usepackage{physics}
\usepackage{comment}
\usepackage{blindtext}
\usepackage{csquotes}
\usepackage[none]{hyphenat}
\DeclareUnicodeCharacter{0304}{ }
\usepackage{rotating}
\usepackage{tabularx}

\usepackage[many]{tcolorbox}
\definecolor{fg}{RGB}{34,139,34}

\renewcommand{\thetable}{\Roman{table}}

\def\figureautorefname~#1\null{Fig.\,#1\null}
\def\equationautorefname~#1\null{Eq.\,(#1)\null}
\def\tableautorefname~#1\null{Tab.\,#1\null}
\definecolor{MyDarkBlue}{rgb}{0.1, 0.1, 0.8} 
\definecolor{MyLightBlue}{rgb}{0.22,0.51,0.9}
\definecolor{MyGreen}{rgb}{0.0, 0.5, 0.0}
\definecolor{BrickRed}{rgb}{0.8, 0.25, 0.33}

\RequirePackage{hyperref}
\hypersetup{colorlinks=true, citecolor=MyGreen,linkcolor=BrickRed, urlcolor=MyLightBlue}
\usepackage{cleveref}
\crefname{equation}{Eq.}{Eqs.} 
\crefrangelabelformat{equation}{(#3#1#4--#5#2#6)}

\title{\bf Quantum Entanglement is Quantum: \\$ZZ$ Production at the LHC}
\author[a]{Dorival Gon\c{c}alves,}
\author[a]{Ajay Kaladharan,}
\author[b]{Frank Krauss,}
\author[a]{Alberto Navarro}
\affiliation[a]{Department of Physics, Oklahoma State University, Stillwater, OK 74078, USA}
\affiliation[b]{Institute for Particle Physics Phenomenology, Durham University, Durham DH1 3LE, UK}
\emailAdd{dorival@okstate.edu}\emailAdd{kaladharan.ajay@okstate.edu}\emailAdd{frank.krauss@durham.ac.uk}\emailAdd{alberto.navarro\_serratos@okstate.edu}

\abstract{
Polarization and spin correlations in diboson systems serve as powerful tools for precision tests and searches for new physics. Recently, interpreting these observables through the lens of quantum information, for instance by examining whether the diboson systems exhibit entanglement, has introduced a compelling new dimension to these studies. We analyze the angular coefficients in the processes $pp\to e^+e^-\mu^+\mu^-$ and $h\to e^+e^-\mu^+\mu^-$, incorporating higher-order QCD and electroweak corrections. Guided by the fundamental properties of the spin density matrix, we assess the stability of the two-qutrit interpretation under radiative effects. For the $pp \to e^+e^-\mu^+\mu^-$ process, NLO QCD corrections preserve the two-qutrit structure but weaken entanglement indicators, an effect that can be partially mitigated by jet binning. In contrast, electroweak corrections introduce non-factorizable contributions that modify the quantum properties of the system. While these effects can be largely depleted by selecting events with a double-resonant $ZZ$ structure, such a kinematic handle is not available for Higgs decays. In the $h \to e^+e^-\mu^+\mu^-$ channel, singly-resonant NLO electroweak corrections substantially distort the angular coefficients, challenging the description of these events as a  two-qutrit system.}

\begin{document}
\maketitle

\begin{sloppypar}

\section{Introduction}
\label{sec:intro}
In recent years, the study of quantum observables in processes involving the production and decay of heavy particles has emerged as a prominent topic in particle physics phenomenology at the Large Hadron Collider (LHC) and future collider experiments.
In particular, entanglement studies have attracted substantial attention, for example in the pair production of top quarks~\cite{Afik:2020onf,Fabbrichesi:2021npl,Severi:2021cnj,Aguilar-Saavedra:2022uye,Afik:2022kwm,Afik:2022dgh,Severi:2022qjy,Aoude:2022imd,Dong:2023xiw,Aguilar-Saavedra:2023hss,Han:2023fci,Sakurai:2023nsc,Cheng:2023qmz,Maltoni:2024tul,Maltoni:2024csn,White:2024nuc,Dong:2024xsg,Dong:2024xsb,Han:2024ugl,Cheng:2025cuv,Afik:2025ejh,Nason:2025hix} or electroweak gauge bosons~\cite{Barr:2021zcp,Aguilar-Saavedra:2022wam,Ashby-Pickering:2022umy,Aoude:2023hxv,Fabbrichesi:2023cev,Fabbrichesi:2023jep,Fabbri:2023ncz,Bernal:2023ruk,Morales:2023gow,Bi:2023uop,Barr:2024djo,Aguilar-Saavedra:2024whi,Subba:2024mnl,Bernal:2024xhm,Sullivan:2024wzl,Aguilar-Saavedra:2024jkj,Grossi:2024jae,Ruzi:2024cbt,Wu:2024ovc,DelGratta:2025qyp,Ding:2025mzj}. The overarching interest in these studies is to establish the presence of quantum correlations, particularly entanglement as one of the hallmarks of quantum physics, at high energy scales. The quantum nature of particle physics is obviously a foundational part of the theoretical framework underpinning calculations in quantum field theory, but experimental confirmation of entanglement at the TeV scale has only recently been achieved, with ATLAS and CMS reporting evidence of spin entanglement in top quark pair production~\cite{ATLAS:2023fsd,CMS-PAS-TOP-23-001,PhysRevD.110.112016}.\footnote{The possibility that entanglement underlies the emergence of key symmetries in particle physics has also been recently explored~\cite{Cervera-Lierta:2017tdt,Beane:2018oxh,Liu:2022grf,Carena:2023vjc,Thaler:2024anb,vonKuk:2025kbv,Carena:2025wyh}.}

In the present study, we focus on the pair production of $Z$ bosons, either directly or as intermediate states in Higgs boson decays including higher-order (i.e.\ quantum) corrections. We employ the framework of quantum tomography to assess the robustness of theoretical observables, their measurement and, ultimately, their interpretation. We use this approach to reconstruct the density matrix $\rho$~\cite{PhysRevLett.83.3103, PhysRevA.64.052312, PhysRevA.66.012303}\footnote{In this work, we perform quantum state tomography of the spin state in the $ZZ$ system. Throughout the discussion, the terms density matrix and entanglement will refer specifically to the spin density matrix and spin entanglement, respectively.}, which encodes the complete information about the corresponding quantum system and possesses key properties that reflect the foundational principles of quantum mechanics:  it is Hermitian, has unit trace, and features real, non-negative eigenvalues that sum to one. We use exactly these fundamental properties to assess the robustness of the two-qutrit interpretation for the diboson pair production in various kinematic regimes.\footnote{For a related study of the $h\to WW^*$ channel, see Ref.~\cite{Goncalves:2025xer}.}

In addition to our discussion of quantum entanglement, we present a detailed analysis of the angular coefficients associated with the processes 
$pp\to e^+e^-\mu^+\mu^-$ and $h\to e^+e^-\mu^+\mu^-$, evaluated at leading order (LO), next-to-leading order (NLO) QCD, and NLO electroweak (EW) accuracy. Although the spin density matrix of the $ZZ$ system may not be well defined in general, depending on the size of some of the higher-order effects, as we investigate in this paper, the angular coefficients remain relevant observables for precision studies and searches for physics beyond the Standard Model (SM)~\cite{ATLAS:2018gqq,Goncalves:2018fvn,Goncalves:2018ptp,ATLAS:2019zrq,CMS:2019nrx,MammenAbraham:2022yxp,Bhardwaj:2023ufl,CMS:2024ony,Carrivale:2025mjy}. It is thus crucial to understand how they are affected by radiative corrections.

This paper is structured as follows.  In \autoref{sec:theory}, we provide a brief review of the theoretical framework involving the formalism of the density matrix, quantum entanglement, and quantum tomography. \autoref{sec:ZZ} addresses quantum tomography in $ZZ$ production at the LHC, with particular emphasis on higher-order QCD and electroweak effects. In \autoref{sec:hZZ}, we focus on $ZZ$ arising from Higgs boson decays and analyze the higher-order EW corrections to the density matrix and its implications for the physics interpretations. Finally, \autoref{sec:summary} is reserved for our summary and outlook.

\section{Theoretical Framework}
\label{sec:theory}

\subsection{Density Matrix and Quantum Entanglement}
\label{sec:entanglement}

The density matrix is a fundamental tool in quantum mechanics, encapsulating all observable information of a quantum system. It generalizes the concept of state vectors, enabling the description of both pure and mixed states. For a pure state $\ket{\Psi}$, the density matrix is given by $\rho = \ket{\Psi}\bra{\Psi}$, while for a mixed state, it is expressed as a convex combination of pure state density matrices
\begin{align}
 \rho = \sum_i p_i \ket{\Psi_i}\bra{\Psi_i},
 \label{eq:density}
\end{align}
where $p_i \geq 0$ are classical probabilities satisfying $\sum_i p_i = 1$. 
From \autoref{eq:density}, a valid density matrix must satisfy the following properties:
\begin{itemize}
    \item The density matrix is Hermitian: $\rho^\dagger = \rho$. 
    \item The trace of the density matrix is equal to 1: $\Tr(\rho) = 1$.   
    \item The density matrix is positive semi-definite, meaning that for any vector  $\ket{\Psi_i}$, $\bra{\Psi_i}\rho\ket{\Psi_i}\ge 0$.
\end{itemize}
These properties imply that the eigenvalues $\lambda_i$ of a generic density matrix must satisfy:
\begin{align}
    0\le\lambda_i\le 1,\quad \text{and} \quad \sum_i \lambda_i=1.
    \label{eq:eigenvalue}
\end{align}

Consider a composite quantum system, consisting of two subsystems $A$ and $B$. The system is defined to be separable if its density matrix $\rho$ can be written as a convex sum
\begin{equation}
    \rho
    \;=\;
    \sum_i p_i \rho_A^i \otimes \rho_B^i,
    \label{eq:seperable}
\end{equation}
where $p_i\geq 0$ and $\sum_i p_i=1$, and  $\rho_A^i$ and $\rho_B^i$ are the density matrices acting on the Hilbert spaces of subsystems $A$ and $B$, respectively. A state that cannot be written in this separable form is named \emph{entangled}~\cite{PhysRevA.40.4277}.

A convenient measure of entanglement is the concurrence $C$, with $C\ne 0$ for entangled systems. 
In the following, we will assume that $A$ and $B$ have equal dimensions, as they will ultimately represent two $Z$ bosons. For such a bipartite system in a pure state $\ket{\Psi}$ with corresponding density matrix $\rho = \ket{\Psi}\bra{\Psi}$, the concurrence $C$ is defined by~\cite{PhysRevA.64.042315, PhysRevLett.80.2245, PhysRevA.54.3824}
\begin{equation}
    \mathcal{C}[\ket{\Psi}]
    \;\equiv\; 
    \sqrt{2\left( 1-\Tr[(\rho_A)^2] \right)}
    \;=\;
    \sqrt{2\left( 1-\Tr[(\rho_B)^2] \right)},
\end{equation}
where $\rho_A=\Tr_B (\rho)$ and $\rho_B=\Tr_A (\rho)$ are the reduced density matrices obtained by tracing over degrees of freedom of subsystems $B$ and $A$, respectively. 
Any mixed state $\rho$ of the bipartite system can be decomposed into a set of pure states $\left\{ \ket{\Psi_i} \right\}$, as denoted in \autoref{eq:density}, and the concurrence is defined as the infimum over all possible decompositions of $\rho$ into pure states $\left\{ p_i,\,\ket{\Psi_i} \right\}$~\cite{Fabbrichesi:2023cev},
\begin{equation}
    \mathcal{C}(\rho)
    \;\equiv\; 
    \inf_{\left\{ \ket{\Psi_i} \right\}}\sum_i p_i \mathcal{C}[\ket{\Psi_i}].
\end{equation}
Evaluating $\mathcal{C}(\rho)$ is generally challenging, with closed-form solutions available only for simple two-level systems.

Fortunately, bounds on $\mathcal{C}(\rho)$ can be employed to estimate entanglement.  
Lower and upper bounds for $\mathcal{C}(\rho)$ are given by~\cite{PhysRevLett.98.140505,Fabbrichesi:2023cev,PhysRevA.78.042308}
\begin{eqnarray}
    (\mathcal{C}(\rho))^2
    &\;\geq\;& 
    2\max\Big\{ 0,\;\; 
    \Tr[\rho^2]-\Tr[(\rho_A)^2],\;\;
    \Tr[\rho^2]-\Tr[(\rho_B)^2] \Big\} 
    \;\equiv\; 
    \mathscr{C}^{2}_{\mathrm{LB}}\,,
    \nonumber\\
    (\mathcal{C}(\rho))^2
    &\;\leq\;& 
    2\min \Big\{ 1-\Tr[(\rho_A)^2],\;\;1-\Tr[(\rho_B)^2] \Big\}
    \;\equiv\;
    \mathscr{C}^{2}_{\mathrm{UB}}\,.
\end{eqnarray}
While a vanishing upper bound, $\mathscr{C}_{\mathrm{UB}}=0$, indicates separable states, a positive lower bound, $\mathscr{C}_{\mathrm{LB}}>0$, confirms the presence of entanglement. In this study, we therefore adopt $\mathscr{C}_{\mathrm{LB}}$ and $\mathscr{C}_{\mathrm{UB}}$ as measures of entanglement for our $ZZ$ systems.

\subsection{Quantum Tomography}
\label{sec:tomography}
Massive vector bosons, such as $Z$ bosons, can be portrayed as qutrits. 
For a system of two such vector bosons, the density matrix $\rho$ acts on the nine-dimensional Hilbert space defined by their spin states. 
This matrix can be parameterized using a basis of irreducible tensor operators~$\left\{ T^{L_1}_{M_1}\otimes T^{L_2}_{M_2}\right\}$ as~\cite{Aguilar-Saavedra:2022wam,Barr:2024djo}
\begin{equation}
    \rho
    \;=\;
    \frac {1}{9}\left( \mathbb{I}_3 \otimes\mathbb{I}_3+A^{1}_{LM} T^{L}_{M}\otimes \mathbb{I}_3+ A^{2}_{LM}\mathbb{I}_3\otimes T^{L}_{M}+ C_{L_1 M_1 L_2 M_2} T^{L_1}_{M_1}\otimes T^{L_2}_{M_2}\right)\,,
\label{eq:rho}
\end{equation}
where $\mathbb{I}_3$ is the identity, and we sum over $L=1,2$ and $-L\leq M \leq L$, with analogous summations for $L_{1,2}$ and $M_{1,2}$. 
The irreducible tensor operators $T^L_M$ satisfy 
\begin{equation}
\Tr\left\{ T^L_M (T^L_M)^\dagger\right\}=3\;\;\;\mbox{\rm and}\;\;\; (T^L_M)^\dagger=(-1)^M T^L_{-M}.
\end{equation}
An explicit representation of the tensor operators is provided in~\autoref{app:TensOp}.
Hermiticity of $\rho$ imposes 
\begin{equation}
A^{1,2}_{LM}=(-1)^M (A^{1,2}_{L,-M})^*\;\;\;\mbox{\rm and}\;\;\; C_{L_1 M_1 L_2 M_2}=(-1)^{M_1+M_2} (C_{L_1, -M_1, L_2, -M_2})^*.
\label{eq:coeff}
\end{equation}
As a result, the density matrix $\rho$ is fully characterized by 80 independent real parameters\footnote{Here, we assume the density matrix to be a general Hermitian matrix with unit trace, which contains 80 independent real parameters. However, depending on the underlying process, many of these parameters are either correlated or vanish, reducing the number of independent parameters. For example, as discussed in~\autoref{subsec:hZZ-LO}, the density matrix for the process $h \to 4\ell$ at LO involves only two independent real parameters.}, which can be extracted from the angular distribution of the vector boson decay products.

Here, we focus on the $ZZ$ system, where each $Z$ boson decays into a pair of charged leptons, $Z\rightarrow \ell^+ \ell^-$. Given the energy scales involved, the charged leptons $(\ell=e,\mu)$ are treated as massless. It is important to emphasize that an off-shell $Z$-boson can be treated as a spin-$1$ particle, because the contribution from the scalar component of its propagator cancels when coupled to massless final-state fermions~\cite{Peskin:1995ev,Aguilar-Saavedra:2022wam}.\footnote{For off-shell gauge bosons with $p_V^2 \neq m_V^2$, the propagator contains a scalar component~\cite{Peskin:1995ev,Korner:2014bca}. When such a boson decays into a fermion $f$, the contribution of this scalar component is proportional to $\mathcal{O}(m_f^2)$. In this work, we focus on $Z$ decays into electron and muon pairs, whose masses are negligible compared to the characteristic energy scale of the process. Therefore, the scalar component of the off-shell gauge boson propagator can be safely neglected.} 
The decay matrix for the $Z$ boson decaying into charged leptons is~\cite{Boudjema:2009fz, Rahaman:2021fcz}
\begin{equation}
\resizebox{\textwidth}{!}{$
    \Gamma
    \;=\;
    \frac 14\begin{pmatrix}
    1+\cos^2 \theta-2\eta_\ell \cos\theta & 
    \frac {1}{\sqrt{2}}\left( \sin 2\theta-2 \eta_\ell\sin\theta \right) e^{i \varphi} & 
    \left( 1-\cos^2 \theta \right) e^{i2 \varphi} \\
    \frac {1}{\sqrt{2}}\left( \sin 2\theta-2 \eta_\ell\sin\theta \right) e^{-i \varphi} & 
    2 \sin^2\theta & 
    -\frac {1}{\sqrt{2}}\left( \sin 2 \theta+2 \eta_\ell \sin \theta \right) e^{i \varphi} \\
    \left( 1-\cos^2 \theta \right) e^{-i2 \varphi} & 
    -\frac {1}{\sqrt{2}}\left( \sin 2 \theta+2 \eta_\ell \sin \theta \right) e^{-i \varphi}  & 
    1+\cos^2\theta + 2 \eta_\ell \cos \theta
\end{pmatrix},
$}
\label{eq:Gamma_mat}
\end{equation}
where $\theta$ and $\varphi$ are respectively the polar and azimuthal  angles of the negatively charged lepton in the rest frame of its parent $Z$ boson, and $\eta_\ell$ is related to the sine of Weinberg angle $\theta_W$ by
\begin{equation}
    \eta_\ell
    \;=\;
    \frac {1-4 \sin^2 \theta_W}{1-4 \sin^2 \theta_W+8 \sin^4 \theta_W}.
\label{eq:etal}
\end{equation}
To evaluate the angles $(\theta, \varphi)$, we adopt the helicity basis~\cite{Bernreuther:2015yna,Aguilar-Saavedra:2022wam}:
\begin{itemize}
    \item $\hat{z}$ is the direction of the $Z$ boson with the highest invariant mass (denoted as $Z_1$) in the $ZZ$ rest frame.
    \item $\hat{x} = \mathrm{sign}(\cos\theta_{\mathrm{CM}})(\hat{p}-\cos\theta_{\mathrm{CM}} \hat{z})/\sin\theta_{\mathrm{CM}}$, where $\hat{p}=(0,0,1)$ 
    and $\cos\theta_{\mathrm{CM}} = \hat{p}\cdot\hat{z}$.
    \item $\hat{y} = \hat{z}\times\hat{x}$.
\end{itemize}

The differential cross-section for the process $ZZ \rightarrow \ell_1^+ \ell_1^- \ell_2^+ \ell_2^-$ can be expressed as~\cite{Aguilar-Saavedra:2022wam}
\begin{equation}
    \frac {1}{\sigma} \frac{d \sigma}{d\Omega_1 d\Omega_2}
    \;=\;
  \left( \frac{3}{4\pi} \right)^2\Tr\left\{ \rho(\Gamma_1 \otimes \Gamma_2)^T \right\},
\label{eq:dsigma1}
\end{equation}
where $\Gamma_j\equiv\Gamma( \theta_j, \varphi_j)$ for $j=1,2$. 
Using~\autoref{eq:rho} and~\autoref{eq:Gamma_mat}, the form in~\autoref{eq:dsigma1} can be further simplified to
\begin{align}
    \frac {1}{\sigma} \frac{d \sigma}{d\Omega_1 d\Omega_2}
    \;=\;
  \frac 1{(4\pi)^2} &\left[ 1+A^1_{LM}B_L Y_L^M(\theta_1, \varphi_1)+ A^2_{LM}B_L Y_L^M(\theta_2, \varphi_2)\right. \nonumber\\
    &\left.+C_{L_1 M_1 L_2 M_2} B_{L_1} B_{L_2} Y_{L_1}^{M_1}(\theta_1, \varphi_1)Y_{L_2}^{M_2}(\theta_2, \varphi_2)\right]\,,
\label{eq:dsigma}
\end{align}
where the summation over the indices $L$ and $M$ is implicit, and we use the following relations:
\begin{eqnarray}
\resizebox{0.98\textwidth}{!}{$
    \Tr\left\{ \mathbb{I}_3 \Gamma^T \right\}
    =
    2\sqrt{\pi}Y_0^0(\theta, \varphi),
    \vphantom{\sqrt{\frac {2\pi}5}}\,
    \Tr\left\{ T^1_M \Gamma^T \right\}
    =
    -\sqrt{2\pi}\eta_\ell Y_1^M(\theta,\varphi),
    \vphantom{\sqrt{\frac {2\pi}5}}\,
    \Tr\left\{ T_M^2 \Gamma^T \right\}
    =
    \sqrt{\frac {2\pi}5}Y_2^M(\theta, \varphi)\,.
    $} \nonumber
\end{eqnarray}
The constants $B_1$ and $B_2$ are given by $B_1=-\sqrt{2 \pi}\eta_\ell$ and $B_2=\sqrt{{2\pi}/{5}}$.

The coefficients $A_{LM}^{1,2}$ and $C_{L_1 M_1 L_2 M_2}$ can be extracted from~\autoref{eq:dsigma} by analyzing the asymmetries~\cite{Aguilar-Saavedra:2017zkn,Rahaman:2018ujg,Rahaman:2021fcz,Rahaman:2019lab,Subba:2024aut} in the angular distributions or applying orthonormality relations of the spherical harmonics~\cite{Baglio:2018rcu,MammenAbraham:2022yxp,Grossi:2024jae}. 
Using the latter, the coefficients $A_{LM}^{1,2}$ and $C_{L_1 M_1 L_2 M_2}$ are given by
\begin{align}
\frac {1}{\sigma} \int \frac{d \sigma}{d\Omega_1 d\Omega_2}Y_L^M(\Omega_i)^{\ast} d \Omega_1 d \Omega_2=& \frac {B_L}{4\pi}A^i_{LM},  \label{eq:Coeffs1}\\
\frac {1}{\sigma} \int \frac{d \sigma}{d\Omega_1 d\Omega_2}Y_{L_1}^{M_1}(\Omega_1)^{\ast} Y_{L_2}^{M_2}(\Omega_2)^{\ast}  d \Omega_1 d \Omega_2=&\frac {B_{L_1} B_{L_2}}{(4\pi)^2}C_{L_1 M_1 L_2 M_2}.\label{eq:Coeffs2}
\end{align}

This procedure guarantees that the reconstructed density matrix is Hermitian $(\rho=\rho^\dagger)$ and has unit trace $(\text{Tr}(\rho)=1)$. However, it does not ensure the positive semi-definiteness condition $(\bra{\Psi} \rho\ket{\Psi}\ge 0,~\forall \ket{\Psi})$. 
The extraction of coefficients using~\autoref{eq:Coeffs1} and~\autoref{eq:Coeffs2} ensures~\autoref{eq:coeff}, preserving the hermicity of the density matrix $\rho$. Additionally, the construction of $\rho$ using~\autoref{eq:rho} satisfies the normalization condition. Nevertheless, the requirement of positive semidefiniteness may still be violated.  
This limitation can arise when the four-lepton final state does not correspond to the two-qutrit $ZZ$ construction described here. Thus, the potential occurrence of a negative eigenvalue would indicate a failure of the procedure.\footnote{The procedure for computing the elements of the density matrix begins by assuming the existence of a two-qutrit $ZZ$ system in~\autoref{eq:rho} and that each $Z$ system can be connected with the decay matrix for two charged leptons in~\autoref{eq:Gamma_mat}. Under these assumptions, the differential cross-section can be written as in~\autoref{eq:dsigma1}. If this yields a nonphysical reconstructed density matrix, specifically, one that is not positive semi-definite, then the procedure is at odds with its own underpinning paradigm and therefore has failed. Such a failure typically signals a breakdown in the validity of at least one of the~\Cref{eq:rho,eq:Gamma_mat,eq:dsigma1}. We investigate these effects in~\autoref{sec:ZZ} and~\autoref{sec:hZZ}.}

A realistic experimental analysis necessarily involves cuts that restrict integration over the full $d\Omega_{1,2}$ phase space, so that the spherical harmonics up to $L=2$ no longer constitute a complete basis. Experimental analyses typically attempt to correct for these effects under the assumption of a complete basis defined over the full phase space. For example, in the ATLAS measurement of angular coefficients in $Z$-boson events~\cite{ATLAS:2016rnf}, the coefficients are extracted from data by fitting templates of the harmonic polynomial terms up to $L=2$ to the reconstructed angular distributions. In contrast to studying the impact of kinematic selections on $d\Omega_{1,2}$ phase space, the present work instead examines the robustness of the complete basis assumption itself, even in the absence of such selections, in light of higher-order effects to the underlying process.

\section{$ZZ$ Production at the LHC}
\label{sec:ZZ}
In this section, we present the results for the production of four charged leptons at the LHC, $pp\to 4\ell$. Our objective is to explore how the various contributions to this process influence the $ZZ$ density matrix, with a focus on their effect for the detection of entanglement. In particular, we analyze the stability and well-defined nature of the quantum observables under the influence of higher-order corrections. In the following, we study the leading order predictions for $pp\to 4\ell$ and evaluate the impact of the NLO QCD and NLO EW effects.

\subsection{Leading Order Production}
\label{subsec:OffshellZ}

\begin{figure}[ht!]
    \centering
    \includegraphics[width=0.8\textwidth]{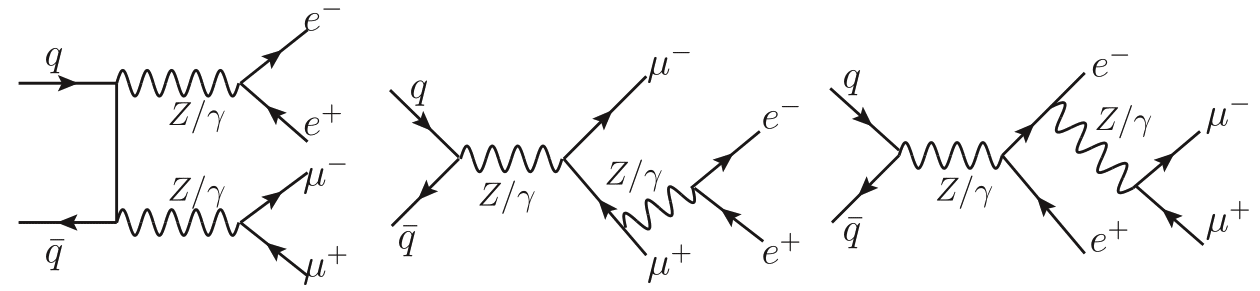}
    \caption{Sample of representative Feynman diagrams for $pp \rightarrow 4\ell$ at leading order.}
    \label{fig:Feyn-DY-LO}
\end{figure}
We start simulating events at parton-level for $pp\to e^{+}e^{-}\mu^{+}\mu^{-}$ at $\sqrt{s}=13$~TeV. To mitigate additional systematic uncertainties arising from identical leptons, we focus on the production of two pairs of leptons from different families. A sample of representative Feynman diagrams contributing to this process at leading order (LO) is depicted in~\autoref{fig:Feyn-DY-LO}. The event generation is performed with \texttt{MadGraph5\_aMC@NLO}~\cite{Alwall:2014hca,Frederix:2018nkq} at LO within the Standard Model (SM), using the \texttt{loop\_qcd\_qed\_sm\_Gmu} UFO model. In our LO calculation, as well as in the NLO QCD and EW studies for $pp\to 4\ell$ presented in the forthcoming sections, we use the following on-shell input parameters:
\begin{align}
    m_h &= 125~\mathrm{GeV}, && \Gamma_h = 4.097~\mathrm{MeV}, \label{eq:mh}\\
    m_t &= 173.2~\mathrm{GeV}, && \Gamma_t =  1.369~\mathrm{GeV}, \label{eq:mt} \\
    m_W &=  80.385~\mathrm{GeV}, && \Gamma_W = 2.085~\mathrm{GeV}, \label{eq:CMparams1}\\
    m_Z &=  91.1876~\mathrm{GeV}, && \Gamma_Z =  2.4952~\mathrm{GeV}, \label{eq:CMparams2}\\
    G_\mu &= 1.1663787\times10^{-5}~\mathrm{GeV}^{-2}. \label{eq:GFparam}&&   
\end{align}
Except for the top quark, all other fermions are treated as massless.  We adopt the complex mass scheme with $G_\mu$ input parameter scheme~\cite{Frederix:2018nkq,Denner:1999gp,Denner:2005fg}. We use the parton distribution functions from \texttt{NNPDF23\_nlo\_as\_0118\_qed}~\cite{Ball:2013hta}, with factorization and renormalization scales set to $\mu_{R}=\mu_{F}=m_Z$, and we impose a minimum invariant mass selection for same-flavor and opposite-sign charged lepton pairs, $m_{\ell^+ \ell^-}>10\, \mathrm{GeV}$, in the event generation.

To investigate the relevant parameter space for entanglement in $ZZ$ production, we generate a second LO sample imposing intermediate diboson production,  $pp\to ZZ\to e^{+}e^{-}\mu^{+}\mu^{-}$, using the same computational framework. The simulation is performed within the double pole approximation (DPA). The corresponding angular coefficients are extracted using~\autoref{eq:Coeffs1} and~\autoref{eq:Coeffs2}. A representative subset of these coefficients for both $pp \to e^{+}e^{-}\mu^{+}\mu^{-}$ and $pp \to ZZ \to e^{+}e^{-}\mu^{+}\mu^{-}$ is presented in~\autoref{tab:coeffsDYLO}. We find good agreement between the two event generation approaches when requiring that the same-flavor and opposite-sign lepton pairs satisfy $|m_{\ell^{+}\ell^{-}} - m_Z| < 10$~GeV.

\begin{table}[htbp]
    \centering
    \begin{tabular}{|c|c|c|}
       \hline Coefficient & $pp\to e^{+}e^{-}\mu^{+}\mu^{-}$ & $pp\to ZZ\to e^{+}e^{-}\mu^{+}\mu^{-}$  \\ \hline
        $A_{2,0}^{1}$ & $0.3313(9)$ & $0.3312(8)$ \\
        $A_{2,0}^{2}$ & $0.3307(9)$ & $0.3325(8)$ \\
        $C_{1,0,1,0}$ & $0.98(1)$ & $0.99(1)$ \\
        $C_{2,0,2,0}$ & $0.234(3)$ & $0.246(3)$ \\
        $C_{2,2,2,2}$ & $-0.398(2)$ & $-0.398(2)$ \\ \hline
    \end{tabular}
    \caption{Angular coefficients at parton level for $pp\to e^{+}e^{-}\mu^{+}\mu^{-}$ and $pp\to ZZ\to e^{+}e^{-}\mu^{+}\mu^{-}$ for $|m_{\ell^{+}\ell^{-}}-m_Z|<10$~GeV. The statistical uncertainty is shown in parentheses.}
    \label{tab:coeffsDYLO}
\end{table}

\begin{figure}[t!]
    \centering
    \includegraphics[width=0.45\textwidth]{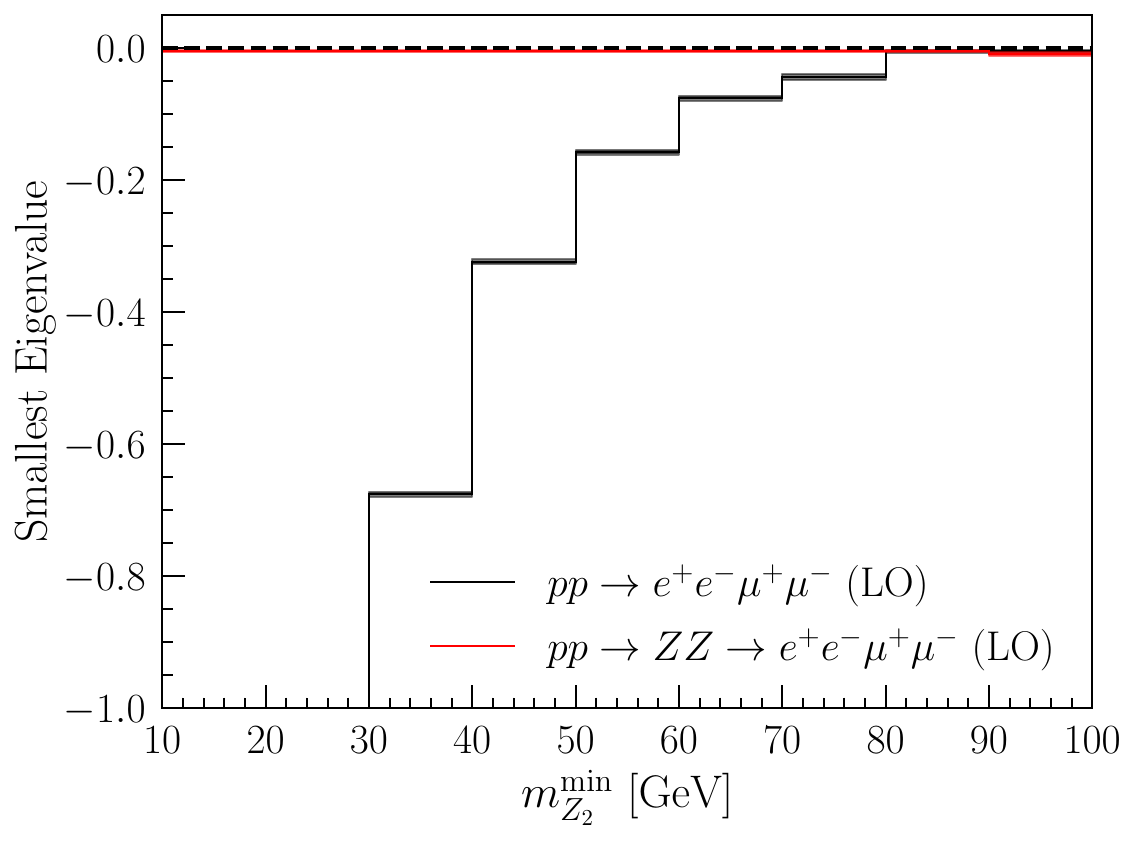}
    \caption{Smallest eigenvalue for the reconstructed diboson density matrix evaluated using the LO $pp\to e^+e^-\mu^+\mu^-$ (black) and $pp\to ZZ\to e^+e^-\mu^+\mu^-$ (red) samples with inclusive results in $m_{4\ell}$ as a function of the lowest reconstructed mass for the same-flavor and opposite-sign dilepton, $m_{Z_2}>m_{Z_2}^\text{min}$. The color band represents the statistical error.}
    \label{fig:eigenvalue}
\end{figure}

We also investigate the two-qutrit interpretation for both event samples by applying the quantum tomography procedure to reconstruct the spin density matrix. 
In~\autoref{fig:eigenvalue}, we illustrate that the interpretation of the $pp \rightarrow e^{+}e^{-}\mu^{+}\mu^{-}$ process in terms of the density matrix for a two-qutrit system can fail, even at LO. The plot shows the smallest eigenvalue for the reconstructed diboson density matrix as a function of the lowest reconstructed mass for the same-flavor and opposite-sign dilepton, $m_{Z_2}$.\footnote{The propagation of uncertainties from the density matrix elements to the eigenvalues and concurrence is performed with the Python package \texttt{pyerrors}~\cite{Joswig:2022qfe}, which is based on the $\Gamma$-method~\cite{Wolff:2003sm, Ramos:2018vgu, Ramos:2020scv}.} 
We observe that when the reconstructed mass for the dilepton system is significantly off-shell from the $Z$ pole, the density matrix displays at least one negative eigenvalue, indicating a breakdown in the two-qutrit physics description. In contrast, for the sample where we impose intermediate $Z$ bosons, $pp\to ZZ\to e^+e^-\mu^+\mu^-$, the smallest eigenvalue of the density matrix remains consistent with zero across the entire kinematic range, indicating a well-defined two-qutrit system. This issue arises for the $pp \rightarrow e^{+}e^{-}\mu^{+}\mu^{-}$ process due to the contributions of Feynman diagrams, such as those shown in~\autoref{fig:Feyn-DY-LO} (central and right diagrams), where a $Z/\gamma$ boson is radiated off a lepton. In such cases, the density matrix formalism for a two-qutrit system, as described in~\autoref{eq:dsigma1}, becomes invalid because the production of the two gauge bosons cannot be treated independently, \emph{i.e.}, one of the gauge bosons is emitted from a decay product of the other. 

\begin{figure}[tb!]
    \centering
    \includegraphics[width=0.45\textwidth]{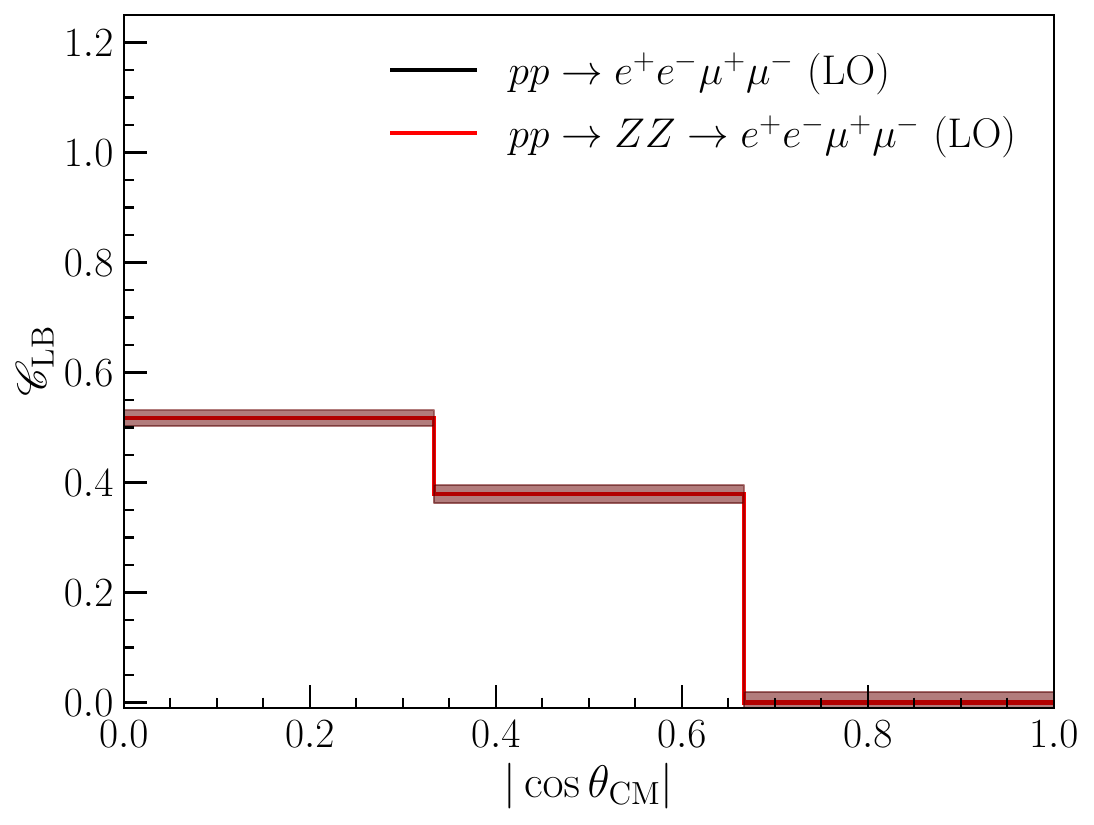}
    \includegraphics[width=0.45\textwidth]{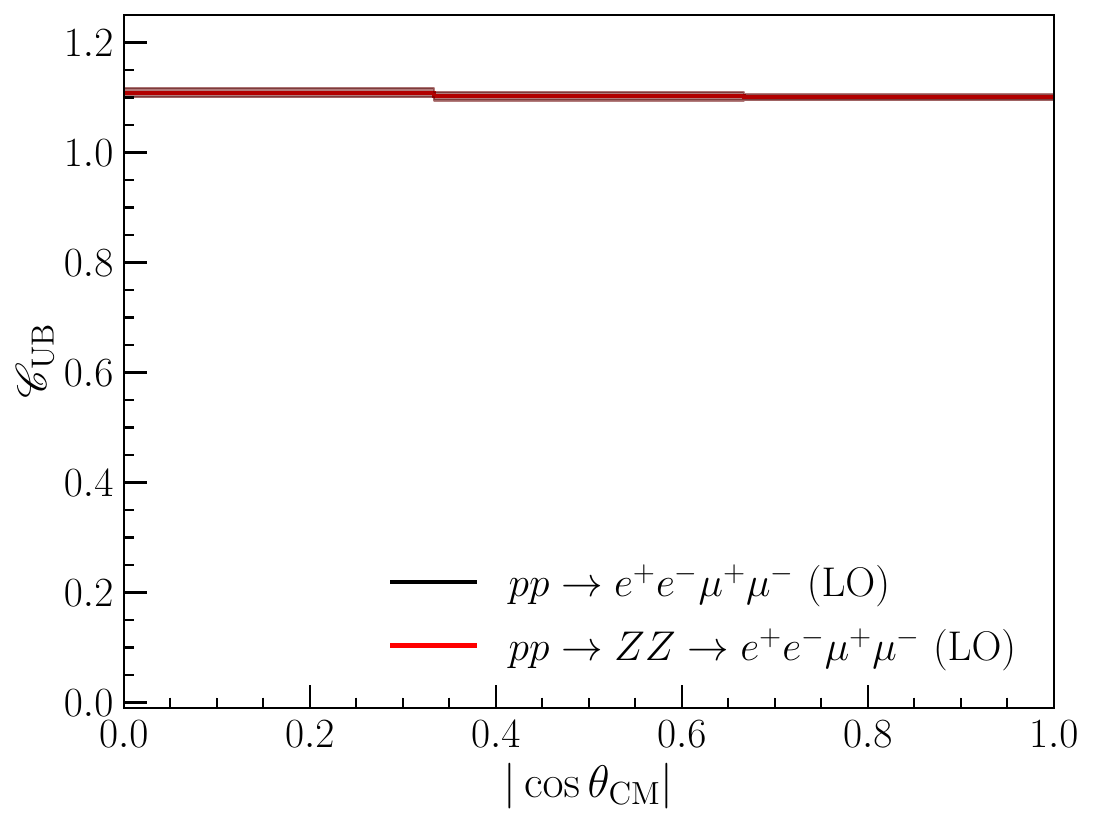}
    \caption{Lower $\mathscr{C}_{\mathrm{LB}}$ (left) and upper $\mathscr{C}_{\mathrm{UB}}$ (right) bounds of the concurrence for $pp\rightarrow e^- e^+ \mu^- \mu^+$ (black) and $pp\rightarrow ZZ\to e^- e^+ \mu^- \mu^+$ (red) at LO as a function of the $\cos\theta_\text{CM}$ with invariant mass selection $81~\text{GeV}<m_{\ell^{+}\ell^{-}}<101$~GeV. The color bands represent the statistical uncertainties.}
    \label{fig:Conc_LO}
\end{figure}

\begin{figure}[b!]
    \centering
    \includegraphics[width=0.45\textwidth]{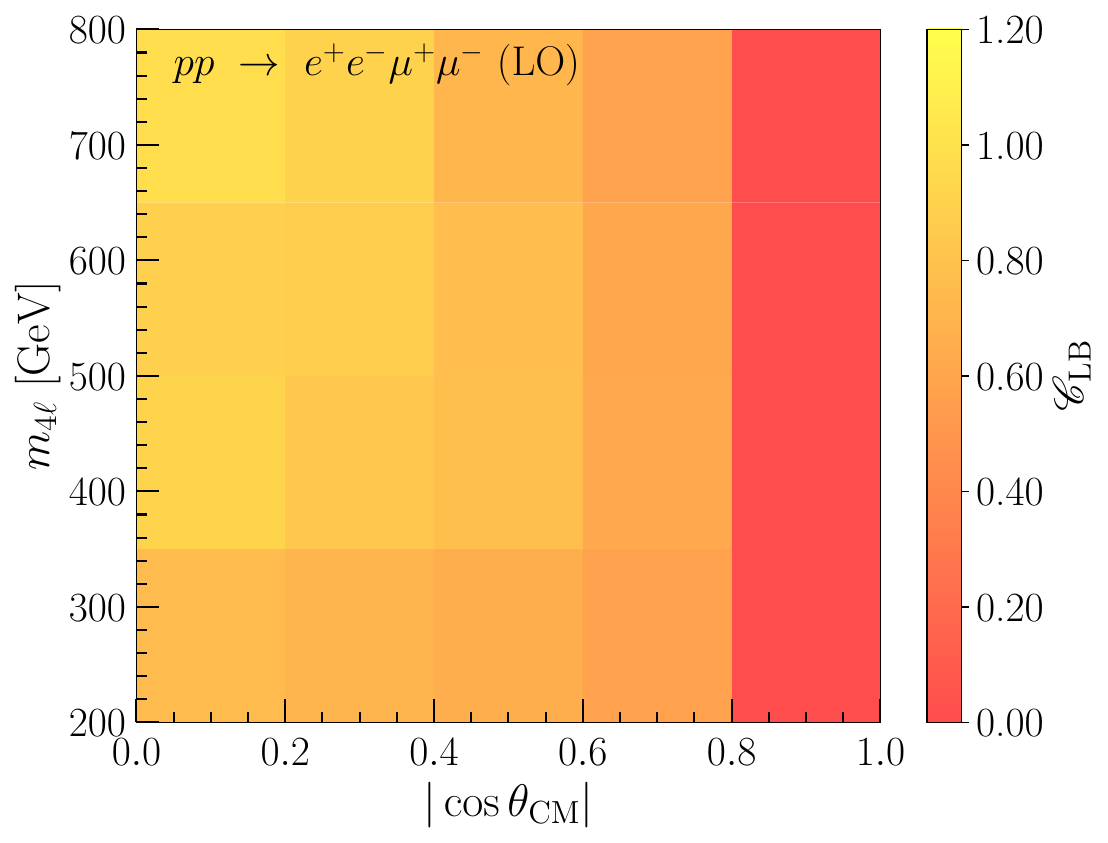}
    \includegraphics[width=0.45\textwidth]{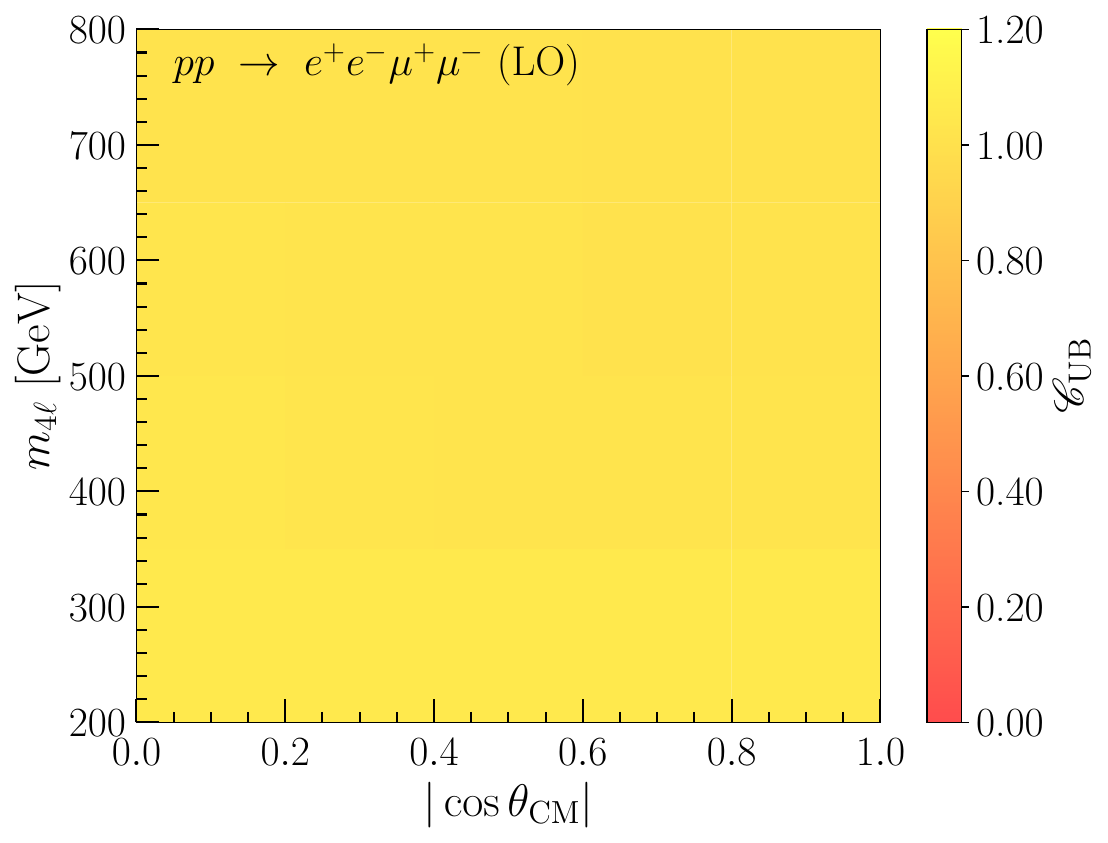}
    \caption{Lower $\mathscr{C}_{\mathrm{LB}}$ (left) and upper $\mathscr{C}_{\mathrm{UB}}$ (right) bounds of the concurrence for the diboson system at LO are shown as functions of the four-lepton invariant mass $m_{4\ell}$ and $|\cos\theta_\text{CM}|$. A narrow  invariant mass window, $81~\text{GeV}<m_{\ell^{+}\ell^{-}}<101$~GeV, is imposed to isolate the $ZZ$ system.}
    \label{fig:Cbounds2D}
\end{figure}

To address this issue, the unwanted LO contributions can be effectively suppressed by imposing a narrow invariant mass window around the $Z$-pole, $81~\text{GeV}<m_{\ell^{+}\ell^{-}}<101$~GeV. Under this condition, the density matrix is well-defined, as observed in~\autoref{fig:eigenvalue}, leading to similar angular coefficients between the two event generation scenarios, as presented in~\autoref{tab:coeffsDYLO}. However, this constraint alone does not yet lead to entanglement in the two-qutrit system, as the lower bound for the concurrence remains at $\mathscr{C}_{\mathrm{LB}}=0$. In~\autoref{fig:Conc_LO}, we present the lower and upper bounds of the concurrence as functions of the cosine of the $Z$ boson scattering angle in the center of mass frame of the $ZZ$ system, $\left| \cos \theta_\text{CM} \right|$, for both $pp \to e^{+}e^{-}\mu^{+}\mu^{-}$  and $pp \to Z Z \to e^+ e^- \mu^+ \mu^-$. While the upper bound remains relatively stable across the phase space, $\mathscr{C}_{\mathrm{UB}}\sim 1.1$, the lower bound increases significantly for central $Z$ bosons, yielding $\mathscr{C}_{\mathrm{LB}}>0$ and signaling the presence of entanglement in this kinematic regime. For comparison, we also include the results from the LO sample $pp \to ZZ \to e^{+}e^{-}\mu^{+}\mu^{-}$, where intermediate $Z$ bosons are explicitly imposed. The concurrence bounds obtained in this scenario show good agreement with those from the inclusive four-lepton production process, confirming that the $ZZ$ density matrix is well-defined at LO after the dilepton invariant mass selections and that it captures the entanglement structure of the system reliably.

To further investigate the kinematic regimes that lead to entanglement in the diboson system, we present in~\autoref{fig:Cbounds2D} the distributions $\mathscr{C}_{\mathrm{LB}}$ (left panel) and $\mathscr{C}_{\mathrm{UB}}$ (right panel) as a function of the four-lepton invariant mass $m_{4 \ell}$ and  $\left| \cos \theta_\text{CM} \right|$. We generally observe a mild increase in $\mathscr{C}_{\mathrm{LB}}$ with larger $m_{4 \ell}$, however, the system displays a strong dependence on the centrality of the $Z$ boson radiation. Entanglement is more pronounced for central $Z$ bosons, corresponding to smaller values of $|\cos\theta_\text{CM}|$ (consistent  with~\autoref{fig:Conc_LO}). In contrast, the upper bound $\mathscr{C}_{\mathrm{UB}}$ remains relatively flat across the phase space, consistently maintaining values above the lower bound $\mathscr{C}_{\mathrm{LB}}$.

\subsection{NLO QCD Effects}
\label{subsec:nlo-qcd}

\begin{figure}[ht!]
    \centering
    \includegraphics[width=0.8\textwidth]{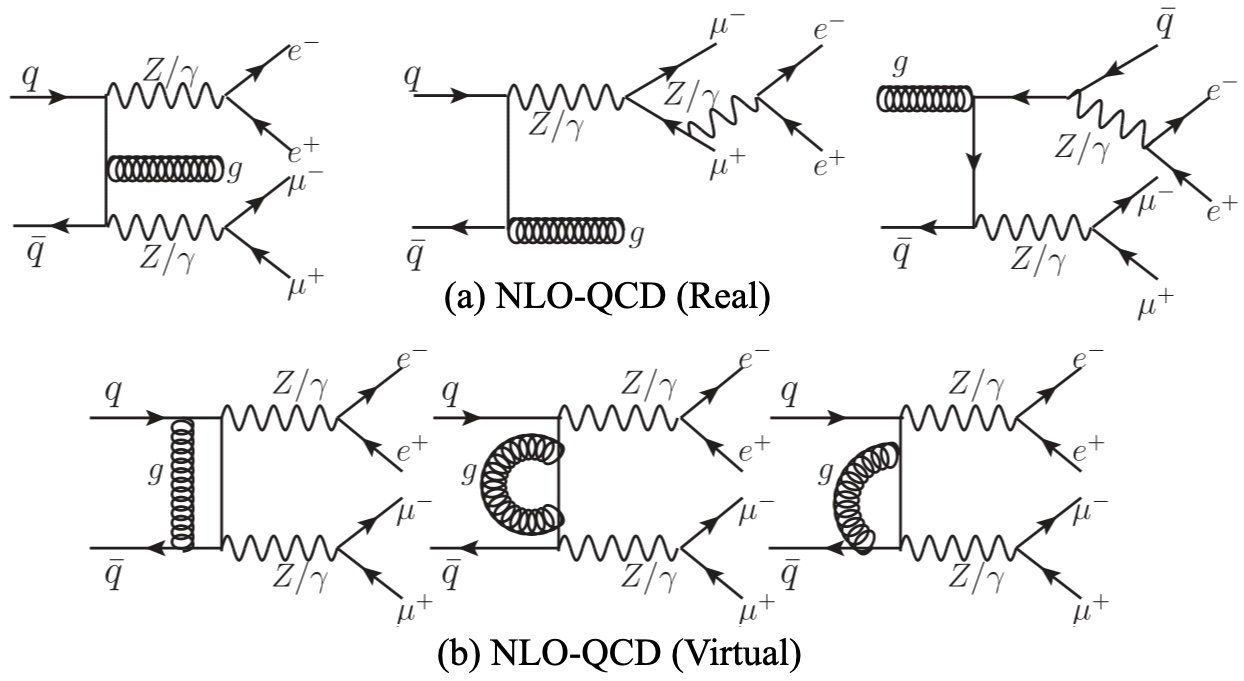}
    \caption{Sample of representative Feynman diagrams for $pp \rightarrow 4\ell$ contributing to the next-to-leading order QCD corrections. Diagrams illustrating real radiation are presented in the top panel (a), while those depicting virtual corrections are shown in the bottom panel (b).}
    \label{fig:Feyn-DY-QCD}
\end{figure}

\begin{table}[htbp]
    \centering
    \begin{tabular}{|c|c|c|c|}
       \hline Coefficient & LO+PS & NLO QCD+PS & NLO/LO  \\ \hline
        $A_{1,0}^{1}$ & $-0.008(2)$ & $-0.035(3)$ & $4.37$ \\
        $A_{2,0}^{1}$ & $0.3224(9)$ & $0.301(1)$ & $0.93$ \\
        $A_{2,1}^{1}$ & $0.0758(6)$ & $0.1023(7)$ & $1.35$\\
        $A_{1,0}^{2}$ & $0.001(1)$ & $0.024(2)$ & $24.0$ \\
        $A_{2,0}^{2}$ & $0.3214(9)$ & $0.299(1)$ & $0.93$ \\
        $A_{2,1}^{2}$ & $0.0987(6)$ & $0.1165(7)$ & $1.18$ \\
        $C_{1,0,1,0}$ & $0.94(1)$ & $0.75(1)$ & $0.80$ \\
        $C_{2,0,2,0}$ & $0.221(2)$ & $0.198(3)$ & $0.89$ \\
        $C_{2,1,2,0}$ & $0.005(2)$ & $0.024(2)$ & $4.80$ \\
        $C_{2,2,2,2}$ & $-0.358(2)$ & $-0.332(2)$ & $0.93$ \\ \hline
    \end{tabular}
    \caption{Angular coefficients for $pp\to e^{+}e^{-}\mu^{+}\mu^{-}$ at LO+PS, NLO QCD+PS,  and their NLO/LO ratio. The selection $|m_{\ell^{+}\ell^{-}} - m_Z| < 10$~GeV is applied to isolate the on-shell $Z$ boson contribution. The statistical uncertainty is shown in parentheses. The imaginary parts of these coefficients are consistent with zero within the uncertainties.}
    \label{tab:coeffsDYLONLOQCD}
\end{table}

\begin{figure}[t!]
    \centering
    \includegraphics[width=0.45\textwidth]{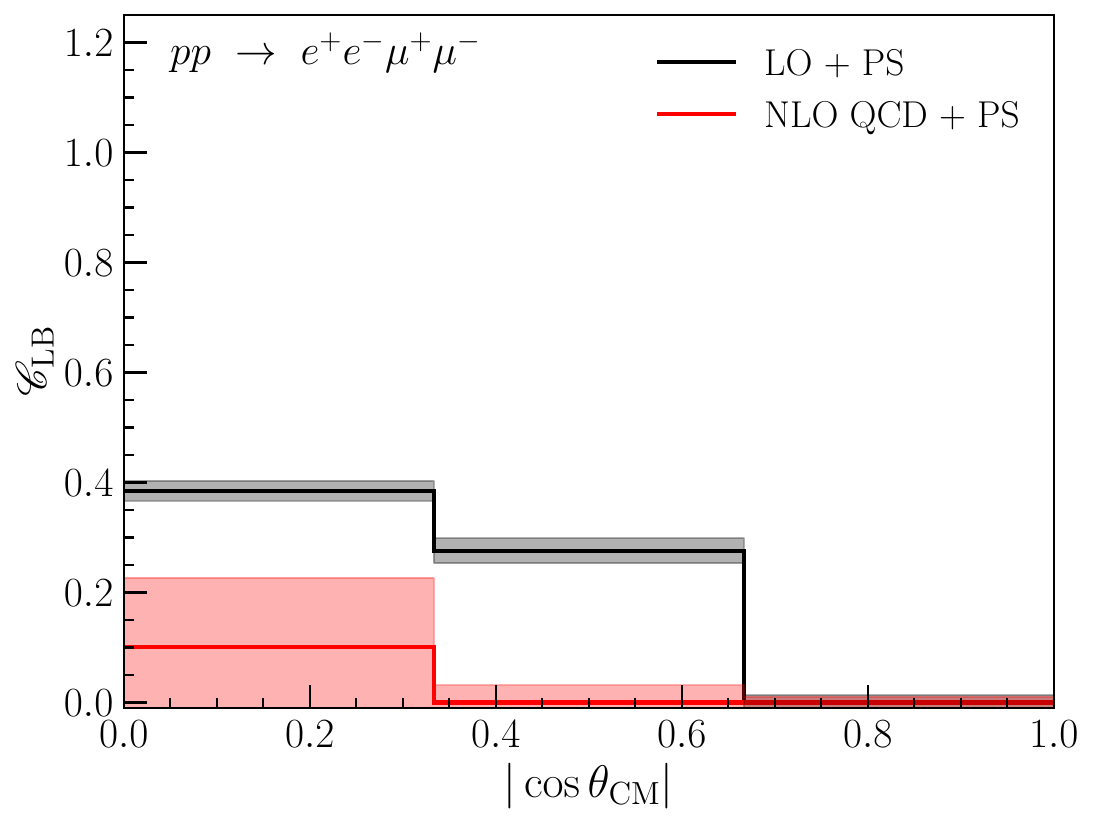}
    \includegraphics[width=0.45\textwidth]{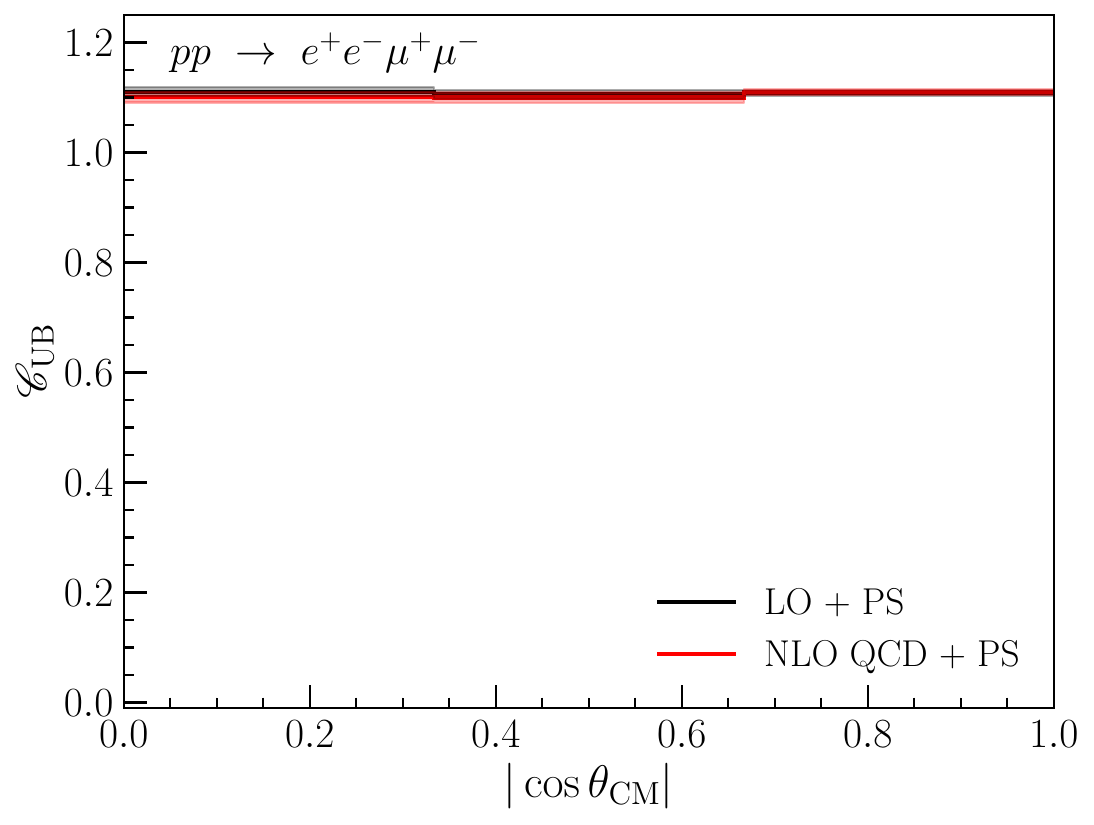} \\
    \includegraphics[width=0.45\textwidth]{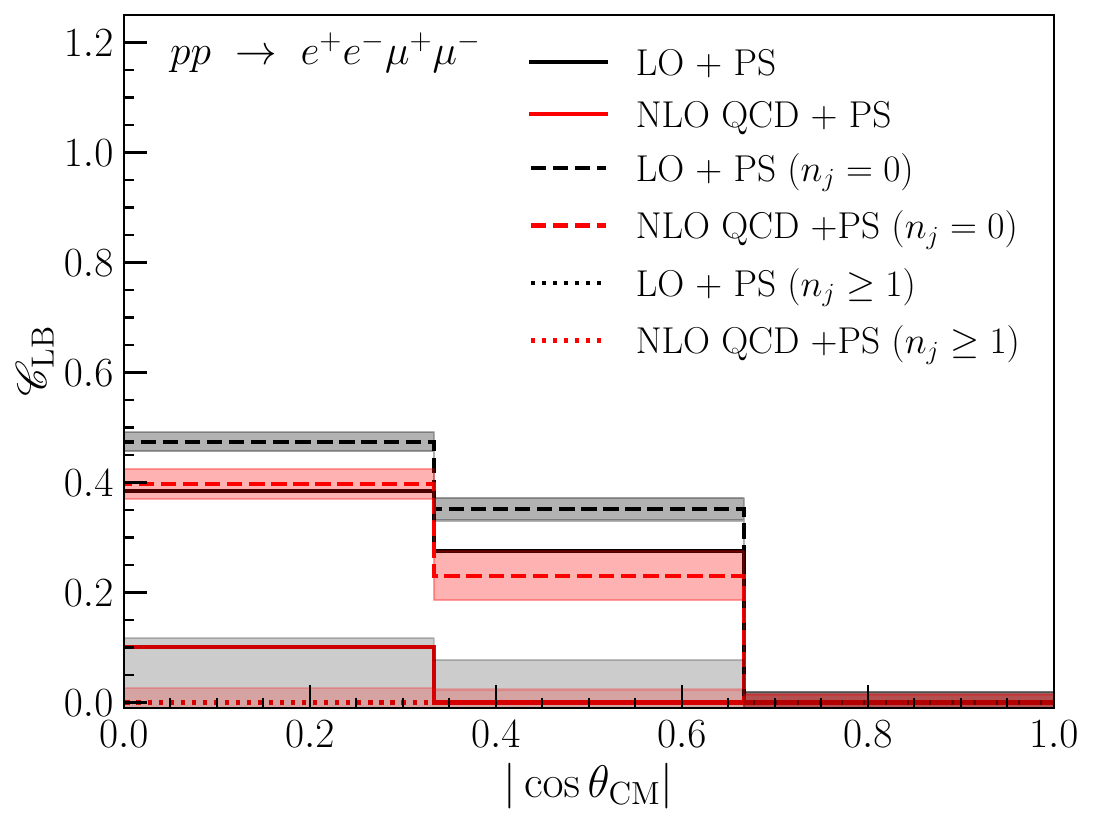}
    \includegraphics[width=0.45\textwidth]{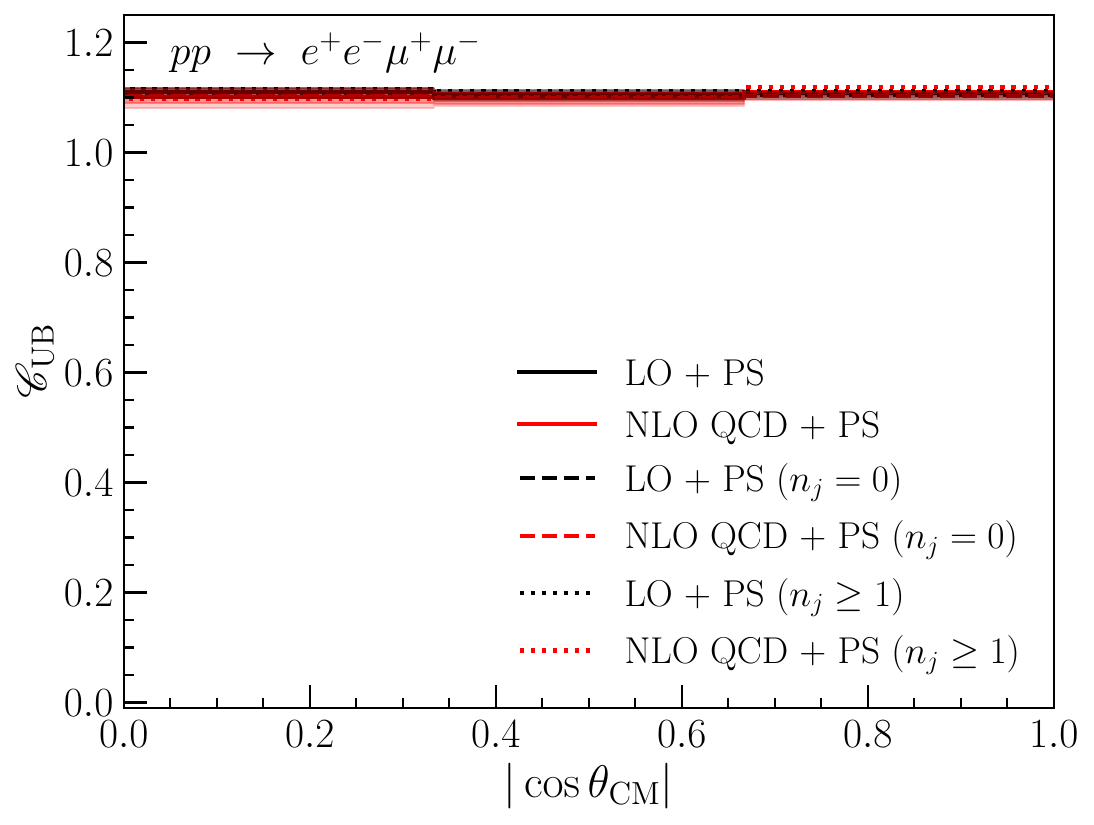}
    \caption{Top panels: Lower $\mathscr{C}_{\mathrm{LB}}$ (left) and upper $\mathscr{C}_{\mathrm{UB}}$ (right) bounds of the concurrence for $pp\rightarrow e^- e^+ \mu^- \mu^+$ as a function of the $\abs{\cos\theta_\text{CM}}$. The black (red) line represents the results at LO+PS (NLO QCD+PS). Bottom panels:  $\mathscr{C}_{\mathrm{LB}}$ and $\mathscr{C}_{\mathrm{UB}}$ at LO+PS (black) and NLO QCD+PS (red). Dashed lines correspond to the $n_{j}=0$ sample, and dotted for $n_{j}\geq 1$. The central values for the inclusive samples are shown with solid lines. The error band accounts for both statistical and theoretical errors. The theoretical error is estimated as the maximum deviation from the central value $\mu_R=\mu_F=m_Z$ when varying the renormalization and factorization scales to $m_Z/2$ and $2m_Z$.}
    \label{fig:CboundsNLO}
\end{figure}

We extend our study of the $pp\to 4\ell$ process by incorporating NLO QCD corrections. While these corrections do not directly affect the $Z$ boson decays to charged leptons, they can modify the helicity structure of the production process, leading to distinct changes in the density matrix and concurrence bounds. A representative sample of Feynman diagrams contributing to this process at NLO QCD is shown in~\autoref{fig:Feyn-DY-QCD}. In particular, real emissions open up the $qg$-initiated channel, see~\autoref{fig:Feyn-DY-QCD} (top right diagram), which alters the helicity configuration of the $Z$ bosons. To investigate these effects, we generate events for $pp\to e^{+}e^{-}\mu^{+}\mu^{-}$ with \texttt{MadGraph5\_aMC@NLO}~\cite{Alwall:2014hca} at NLO QCD matched to Parton Shower (PS) with \texttt{Pythia8.3}~\cite{Bierlich:2022pfr}. The loop-induced $gg$ channel provides an additional contribution to $ZZ$ production, which, although identified as NNLO in QCD, leads to a $\sim 15\%$ increase in the total cross section~\cite{Buschmann:2014sia,Cascioli:2014yka}. These NNLO contributions are not included in this study. As in the parton level analysis,  we impose an invariant mass window around the $Z$ pole, $81~\text{GeV}<m_{\ell^{+}\ell^{-}}<101$~GeV, to isolate the $ZZ$ system  (see~\autoref{subsec:OffshellZ}). For simplicity, hadronization effects are turned off. Additionally, we generate a comparison sample in which LO events are matched with PS. Jets are defined using the anti-$k_T$ algorithm~\cite{Cacciari:2008gp} with radius parameter $R=0.4$, requiring $p_{Tj}>30$~GeV and $\abs{\eta_j}<3$.

We first study the impact of the NLO QCD effects on the angular coefficients. In~\autoref{tab:coeffsDYLONLOQCD}, we present a subset of the angular coefficients for $pp\to e^{+}e^{-}\mu^{+}\mu^{-}$ at LO+PS and NLO QCD+PS. The higher-order corrections display sizable effects, significantly shifting the angular coefficients~\cite{Grossi:2024jae}.\footnote{For a comparison of NLO fixed-order and NLO+PS predictions in the $pp\to 4\ell$ channel, see~\autoref{app:compPS}.} These modifications suggest that quantum entanglement is likely sensitive to NLO QCD effects. To quantify this, we analyze the concurrence bounds in~\autoref{fig:CboundsNLO}, where the upper (top right panel) and lower (top left panel) bounds of the concurrence are shown as functions of $\abs{\cos\theta_\text{CM}}$ for both the LO+PS and NLO QCD+PS samples. 
 While the upper bound $\mathscr{C}_{\mathrm{UB}}$ remains stable across all studied scenarios, the lower bound $\mathscr{C}_{\mathrm{LB}}$ shows significant differences. More concretely, the NLO QCD+PS effects (red solid line) substantially reduce the lower bound for entanglement $\mathscr{C}_{\mathrm{LB}}$ compared to the LO+PS results (black solid line). The lower bound decreases across the entire phase space, becoming zero in some regions where it was positive at LO. To explore this further, we show the concurrence bounds in the bottom panels of \autoref{fig:CboundsNLO}, separating the NLO QCD+PS events into distinct jet multiplicities: $n_j=0$ (red dashed lines) and $n_j\ge 1$ (red dotted lines). For the zero-jet sample, $\mathscr{C}_{\mathrm{LB}}$ at NLO QCD is closer to the LO result,  whereas the $n_j\ge 1$ sample leads to null $\mathscr{C}_{\mathrm{LB}}$, indicating that the $ZZ$ system is no longer entangled. These findings underscore the importance of the higher-order QCD corrections in determining the entanglement behavior of the system, which is crucial for designing tailored experimental analyses.\footnote{Our analysis demonstrates for the first time that NLO QCD corrections in $pp \to 4\ell$ have a strong impact on quantum entanglement: they significantly deplete the concurrence lower bound, such that an entangled two-qutrit state can no longer be established within uncertainties in relevant regions of the parameter space. We further show that entanglement can be restored through a jet-binned analysis, see~\autoref{fig:CboundsNLO}.}

\subsection{NLO EW Effects}
\label{subsec:nlo-ew}

\begin{figure}[ht!]
    \centering
    \includegraphics[width=0.8\textwidth]{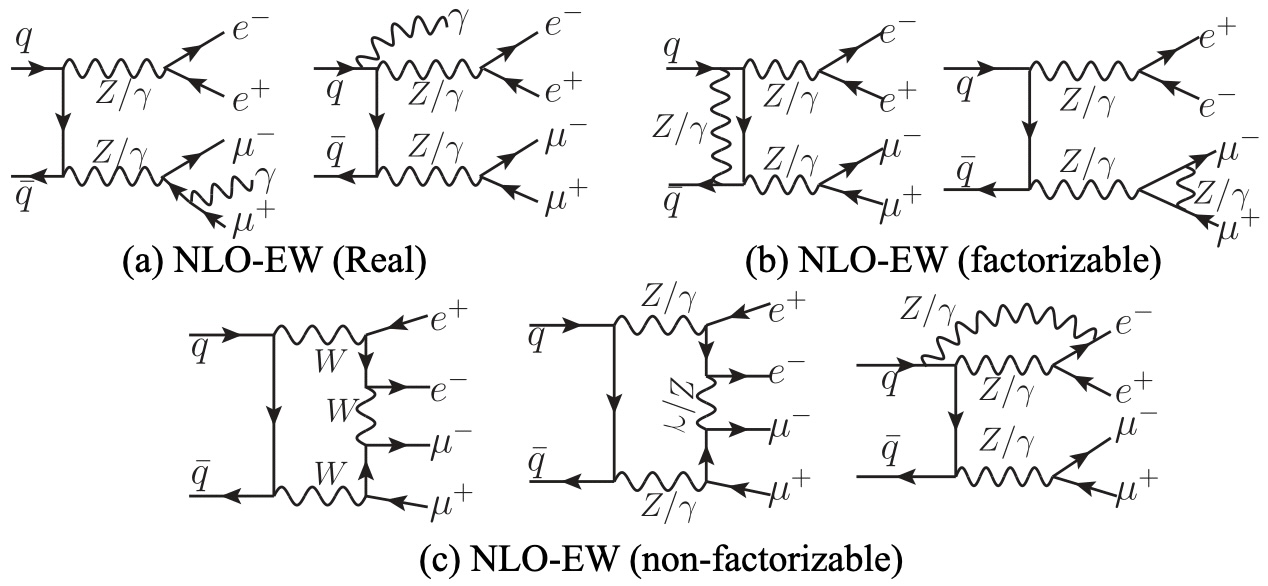}
    \caption{Representative sample of Feynman diagrams for the NLO EW corrections to $pp \rightarrow 4\ell$.}
    \label{fig:Feyn-DY-EW}
\end{figure}

We conclude our analysis of $pp\to 4\ell$ by studying the impact of the NLO EW corrections. A representative set of Feynman diagrams illustrating these contributions is shown in~\autoref{fig:Feyn-DY-EW}. Unlike NLO QCD corrections, which exclusively modify the helicity structure of the diboson production while keeping the decay process unaffected, NLO EW corrections also affect the final-state leptons~\cite{Grossi:2024jae, DelGratta:2025qyp}. These additional contributions could, in principle, disrupt the phenomenological formalism introduced in~\autoref{sec:tomography} if they become significant. For example, the decay of the two $Z$ bosons may no longer factorize, as illustrated by the virtual corrections shown in~\autoref{fig:Feyn-DY-EW}~(c). Additionally, real photon emissions from final-state leptons, as illustrated in~\autoref{fig:Feyn-DY-EW}~(a), could distort the structure of the decay matrix in~\autoref{eq:Gamma_mat}. Virtual corrections to the $Z\ell^+\ell^-$ vertex, shown in the second diagram of~\autoref{fig:Feyn-DY-EW}~(b), could also modify the spin analyzing factor $\eta_\ell$ from its LO value. These contributions could invalidate the formalism outlined in~\autoref{sec:tomography}, making the density matrix interpretation of the diboson system no longer applicable.

To investigate the numerical relevance of these new contributions, we simulate events for $pp\to e^+ e^-\mu^+\mu^-$ at NLO EW accuracy, matched to QCD+QED parton-shower (PS) effects, using the publicly available \texttt{POWHEG-BOX-RES} Monte Carlo~\cite{Chiesa:2020ttl,Alioli:2010xd, Frixione:2007vw, Nason:2004rx, Jezo:2015aia}. In this setup, the Born, virtual, and real matrix elements are computed using \texttt{RECOLA2}~\cite{Denner:2017vms, Denner:2017wsf, Actis:2012qn, Actis:2016mpe}. The \texttt{COLLIER} library is used to reduce the tensor integrals and to evaluate the scalar integrals from the one-loop diagrams~\cite{Denner:2014gla}. The input parameters are identical to those described in~\cref{eq:mh,eq:mt,eq:CMparams1,eq:CMparams2,eq:GFparam}.
We dress the charged leptons with photons within $\Delta R( \ell, \gamma)<0.1$ and impose the invariant mass window of $81~\mathrm{GeV}<m_{\ell^{+}\ell^{-}} < 101~\mathrm{GeV}$. 

Although the extraction of $A_{LM}^{i}$ and $C_{L_1M_1L_2M_2}$ with $L=1$ using~\cref{eq:Coeffs1,eq:Coeffs2} displays sensitivity to the weak mixing angle, $\sin\theta_W$, through the presence of $\eta_\ell$ on the right-hand side of the equations, the angular coefficients $A_{LM}^i$ and $C_{L_1,M_1,L_2,M_2}$ parameterize the production density matrix of the $ZZ$ system and must be independent of $\sin\theta_W$. This apparent dependence cancels out from both sides of~\cref{eq:Coeffs1,eq:Coeffs2} if the values of the weak mixing angle used in the differential distribution, $d\sigma/d\Omega_1 d\Omega_2$, and in the quantum tomography are equal. This suggests that incorporating higher-order EW corrections to the weak mixing angle may be necessary to improve the accuracy of these coefficients when applying quantum tomographic analysis to NLO EW event samples. To evaluate the impact of such corrections, we perform the quantum tomography using~\cref{eq:Coeffs1,eq:Coeffs2} under two different prescriptions for the weak mixing angle:{\footnote{The quantum tomography analyses are performed on the same event sample using two prescriptions for the weak mixing angle, $\sin^2 \theta_W^{\rm LO}$ and $\sin^2 \theta_{\rm eff}^\ell$.}}
\begin{itemize}
    \item As a baseline, we adopt the tree level relation, valid to all orders in the on-shell renormalization scheme, relating the weak mixing angle to the $W$ and $Z$ boson masses:
    \begin{align}
        \sin^2\theta_W^\text{LO}=1-\frac{m_W^2}{m_Z^2}.
        \label{eq:sw2LO}
    \end{align}
    \item To account for electroweak radiative corrections, we alternatively use the effective mixing angle, $\sin^2\theta_\text{eff}^{\ell}$. This quantity absorbs loop corrections and is directly defined in terms of the relative strengths of the axial and vector effective couplings, $a_\ell$ and $v_\ell$, of the $Z$ boson to leptons $\ell$. Parametrizing the $Z$ boson-lepton vertex as $\Gamma_{Z\ell^+\ell^-}=i\bar{\ell}\gamma^\mu(v_\ell+a_\ell \gamma_5)\ell Z_\mu$, the effective mixing angle reads~\cite{Sirlin:2012mh,Awramik:2006uz,Chiesa:2019nqb}: 
    \begin{align}
        \sin^2\theta_\text{eff}^{\ell}=\frac{1}{4}\left(1+\mathrm{Re}\frac{v_\ell}{a_\ell}\right).
        \label{eq:sw2eff}
    \end{align}
    This definition captures universal corrections arising from propagator self-energies and non-universal (flavor-specific) vertex corrections, making it more suitable for comparisons with experimentally extracted values and Monte Carlo event samples that include NLO electroweak effects~\cite{ALEPH:2005ab,ATLAS:2018gqq,CMS:2024ony,LHCb:2024ygc,Biekotter:2023vbh}. At tree level, \autoref{eq:sw2LO} and \autoref{eq:sw2eff} coincide but can differ once radiative corrections are included. In our analysis, $\sin^2\theta_\text{eff}^{\ell}$ is computed at one-loop level using the \texttt{GRIFFIN} package~\cite{Chen:2022dow}, adopting the $(G_\mu, m_W, m_Z)$ input scheme with parameters specified in~\cref{eq:mh,eq:mt,eq:CMparams1,eq:CMparams2,eq:GFparam}.
\end{itemize}

\begin{table}[t!]
\centering
\begin{tabular}{
|c|c|c|c|}
\hline
\multicolumn{1}{|c|}{} & \multicolumn{1}{c|}{LO} & \multicolumn{1}{c|}{NLO} & \multicolumn{1}{c|}{NLO/LO} \\
\hline
$\sin^2 \theta_W$& $0.222897$ &  $0.232289$ & 1.04 \\
\hline
$\eta_\ell$& $ 0.214303$&  $0.140981$ & 0.66\\
\hline
\end{tabular}  
\caption{Weak mixing angle $\sin^2 \theta_W$ at LO ($\sin^2\theta_W^\text{LO}$) and NLO ($\sin^2\theta_\text{eff}^\ell$), and their NLO/LO ratio. The corresponding spin analyzing powers $\eta_\ell$ are also presented. The LO and NLO weak mixing angles are defined in~\cref{eq:sw2LO,eq:sw2eff}, respectively. The effective weak mixing angle is calculated at one-loop with the \texttt{GRIFFIN} package~\cite{Chen:2022dow}, using the $(G_\mu, m_W, m_Z)$ input scheme with parameters specified in~\cref{eq:mh,eq:mt,eq:CMparams1,eq:CMparams2,eq:GFparam}.} 
\label{tab:sinw2}
\end{table}

\begin{table}[t!]
    \centering
    \begin{tabular}{|c|c|c|c}
       \hline Coefficient & \thead{NLO EW+PS \\ $\sin^2\theta_W^\text{LO}$} & \thead{NLO EW+PS
       \\ $\sin^2\theta_\text{eff}^\ell$}  \\ \hline
        $A_{1,0}^{1}$ & $0.002(2)$ & $0.004(3)$ \\
        $A_{2,0}^{1}$ & $0.325(1)$  & $0.325(1)$ \\
        $A_{2,1}^{1}$ & $0.0949(7)$ & $0.0949(7)$ \\
        $A_{1,0}^{2}$ & $-0.002(3)$ & $-0.003(3)$ \\
        $A_{2,0}^{2}$ & $0.338(1)$ & $0.338(1)$ \\
        $A_{2,1}^{2}$ & $0.1018(7)$ & $0.1018(7)$ \\
        $C_{1,0,1,0}$ & $0.27(1)$ & $0.62(3)$ \\
        $C_{1,-1,1,1}$  & $-0.055(9)$   & $-0.13(2)$ \\
        $C_{1,-1,1,-1}$ & $-0.057(9)$ &  $-0.13(2)$ \\
        $C_{2,0,2,0}$ & $0.234(3)$ & $0.234(3)$ \\
        $C_{2,1,2,0}$ & $-0.163(2)$ & $-0.163(2)$  \\
        $C_{2,2,2,2}$ & $-0.381(2)$ & $-0.381(2)$ \\ \hline
    \end{tabular}
    \caption{Angular coefficients for $pp\to e^{+}e^{-}\mu^{+}\mu^{-}$ at NLO EW+PS, computed using $\sin^2\theta_W^\text{LO}$ and $\sin^2\theta_\text{eff}^\ell$ for the quantum tomography study, with $|m_{\ell^{+}\ell^{-}}-m_Z|<10$~GeV. The statistical uncertainty is shown in parentheses. The imaginary parts of these coefficients are consistent with zero within the uncertainties.}
    \label{tab:coeffsDYLONLOEW}
\end{table}

\begin{figure}[tb!]
    \centering
    \includegraphics[width=0.48\textwidth]{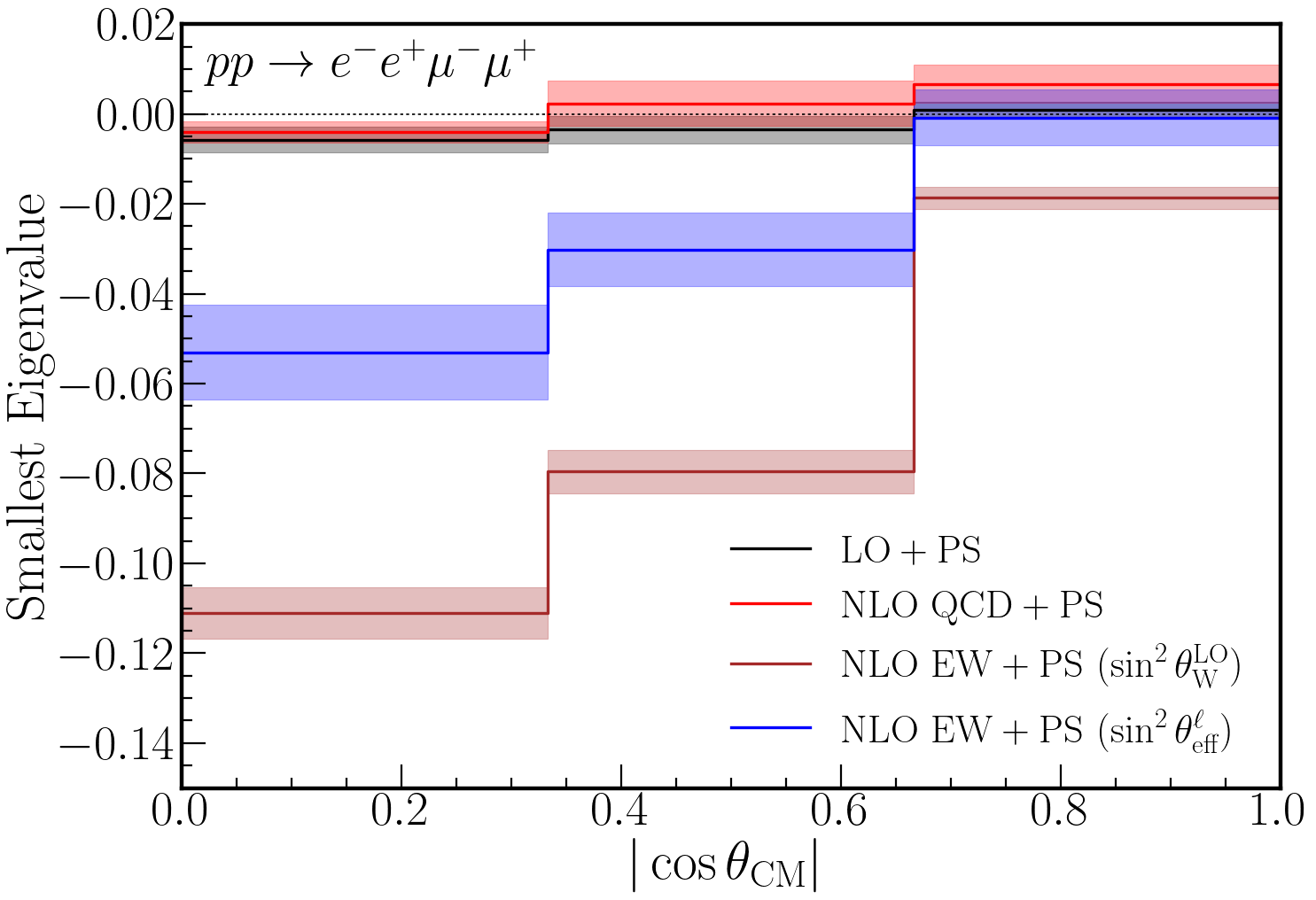}
    \includegraphics[width=0.48\textwidth]{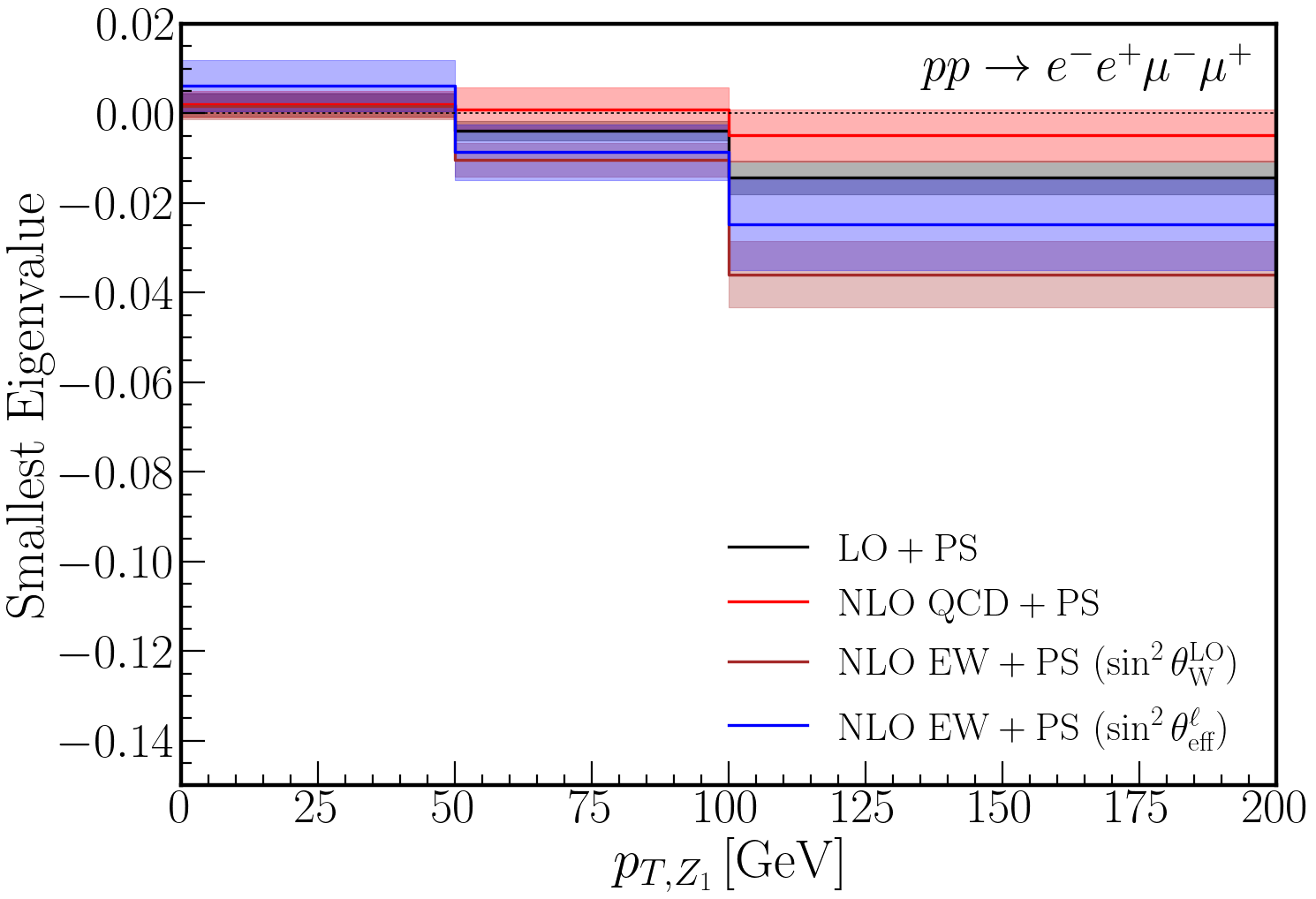}
    \caption{Smallest eigenvalue of the density matrix for the reconstructed diboson system as a function of $|\cos \theta_{\rm{CM}}|$ (left panel) and $p_{T,Z_{1}}$ (right panel) for the process $pp \rightarrow e^+e^-\mu^+\mu^-$ at LO (black), NLO QCD+PS (red), NLO EW+PS using $\sin^2\theta_W^\text{LO}$ for the quantum tomography analysis (brown), and NLO EW+PS with $\sin^2\theta_\text{eff}^\ell$ (blue).  The error bands account for both statistical and theoretical uncertainties. The theoretical error is estimated as the maximum deviation from the central value $\mu_R=\mu_F=m_Z$ when varying the renormalization and factorization scales to $m_Z/2$ and $2m_Z$.}
    \label{fig:SMEIG_PP_4l_NLOEW}
\end{figure}

While higher-order corrections to the weak mixing angle are relatively mild, resulting in a $\sim 4\%$ shift when moving from $\sin^2\theta_W^\text{LO}$ to $\sin^2\theta_\text{eff}^\ell$, we find that the corresponding higher-order effects are augmented in the spin analyzing power, $\eta_\ell^\text{NLO}/\eta_\ell^\text{LO} \approx 0.66$, see~\autoref{tab:sinw2} for more details. These sizable effects suggest that relying on the LO weak mixing angle could invalidate the quantum tomographic reconstruction for the NLO EW samples. We illustrate these effects by computing the angular coefficients with  $\sin^2\theta_W^\text{LO}$ and $\sin^2\theta_\text{eff}^\ell$. The corresponding results are presented in~\autoref{tab:coeffsDYLONLOEW}. As expected, the $L=1$ coefficients display significant shifts when replacing the LO weak mixing angle with its effective counterpart. Furthermore, we observe substantial NLO EW corrections when comparing these results with those shown in~\autoref{tab:coeffsDYLONLOQCD}. In particular, the coefficients $C_{2,1,2,0}$ and $C_{1,0,1,0}$ changed by factors of $-32$ and $0.66$, respectively, when moving from LO+PS to NLO EW+PS with effective weak mixing angle.

We investigate the phase space regions in which the NLO EW corrections may disrupt the validity of the density matrix for a two-qutrit system. For fully inclusive events the density matrix remains well-defined with smallest eigenvalue found to be 0.005(3) and  0.004(4) when using $\sin^2 \theta_W^{\rm{LO}}$ and $\sin^2 \theta_{\rm{eff}}^{\ell}$, respectively.
To assess the robustness of the density matrix across different kinematic regimes, we present in~\autoref{fig:SMEIG_PP_4l_NLOEW} the smallest eigenvalue of the diboson spin density matrix at LO, NLO QCD, and NLO EW accuracy. The NLO EW results are shown for two definitions of the weak mixing angle. In the left panel of~\autoref{fig:SMEIG_PP_4l_NLOEW}, we present the lowest eigenvalue of the diboson density matrix as a function of $|\cos \theta_{\rm{CM}}|$, which is particularly relevant since entanglement typically arises for lower values of $|\cos \theta_{\rm{CM}}|$, as shown with the LO and NLO QCD samples, see~\autoref{subsec:OffshellZ} and~\autoref{subsec:nlo-qcd}, respectively. While the density matrix is well defined for the LO and NLO QCD samples, considering the uncertainties, the NLO EW sample displays relatively large negative eigenvalues, particularly near $|\cos \theta_{\rm{CM}}|\sim 0$.
Remarkably, the smallest eigenvalue moves closer to zero when $\sin^2 \theta_\text{eff}^\ell$ is used in place of $\sin^2 \theta_W^{\rm{LO}}$. This behavior indicates that some NLO EW effects, which disrupt the density matrix formalism for the two-qutrit system presented in~\autoref{sec:tomography}, can be partially mitigated by incorporating one-loop corrections to the weak mixing angle through $\sin^2 \theta_\text{eff}^\ell$. 

\begin{figure}[!tb]
    \centering
    \includegraphics[width=0.48\textwidth]{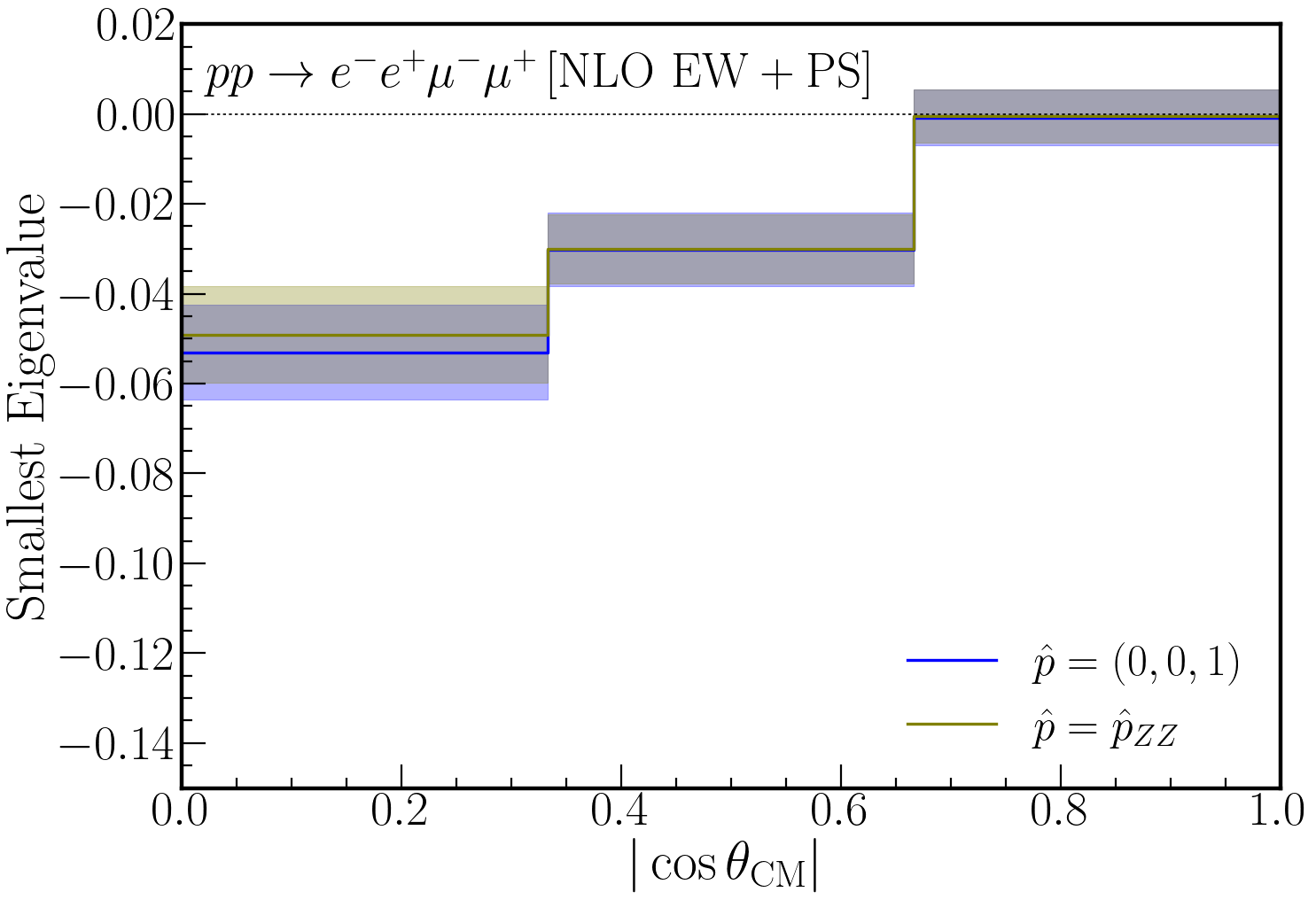}
    \includegraphics[width=0.48\textwidth]{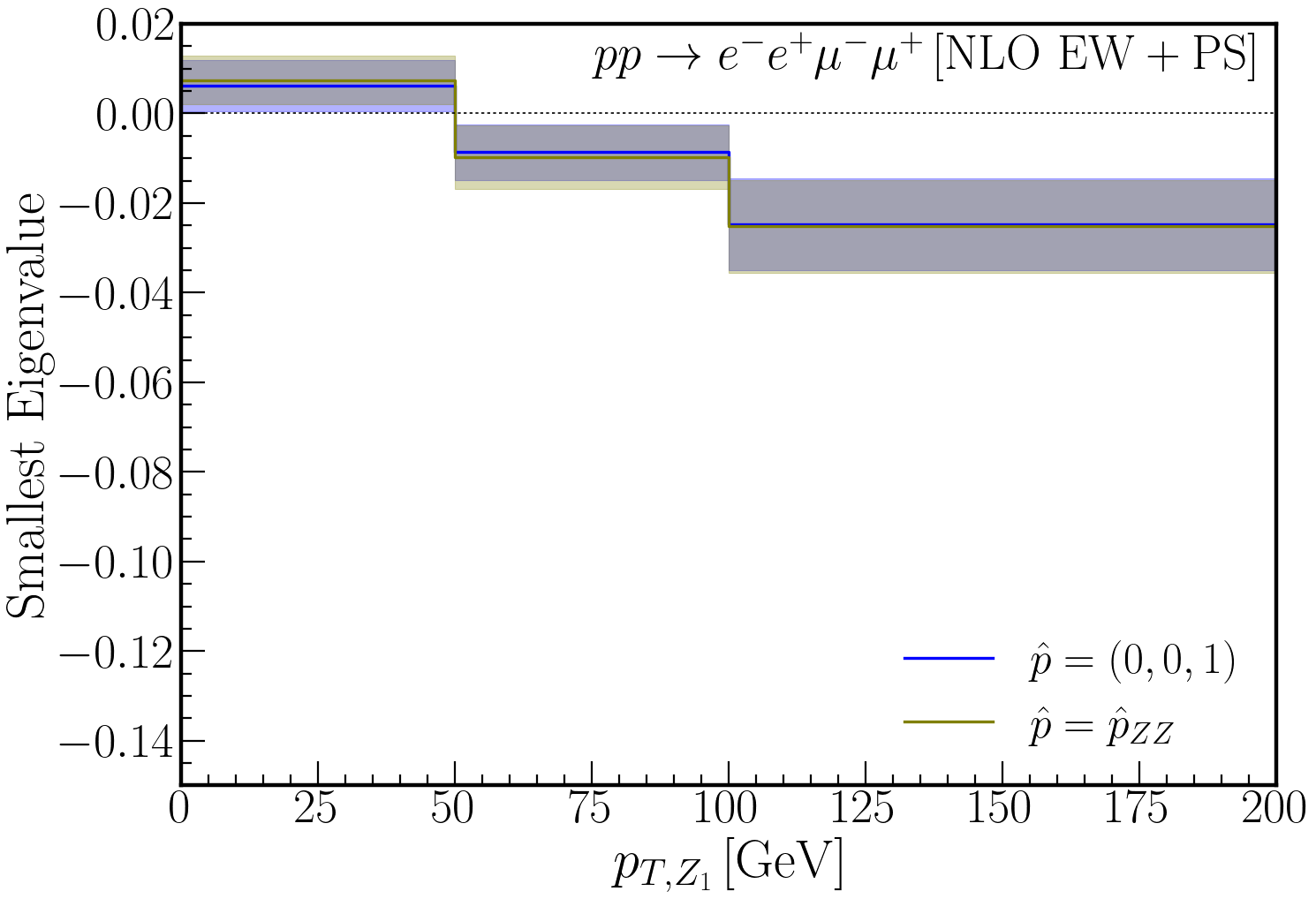}
    \caption{Smallest eigenvalue of the density matrix evaluated in two different reference frames: the helicity frame described in~\autoref{sec:tomography}, labeled as $\hat{p}=(0,0,1)$, and an alternative frame where the reference axis is aligned with the diboson system direction in the laboratory frame, $\hat{p}=\hat{p}_{ZZ}$. The results were derived with NLO EW+PS sample using the quantum tomography approach with $\sin^2\theta_\text{eff}^\ell$.}
    \label{fig:SMBPlots}
\end{figure}

Since the values of the coefficients $A^i_{LM}$ and $C_{L_1M_1L_2M_2}$ depend on the choice of reference frame, it is important to probe whether the results change significantly when moving from the helicity frame, described in~\autoref{sec:tomography}, to an alternative frame. One such alternative frame, adopted in Ref.~\cite{Grossi:2024jae}, involves replacing the reference axis $\hat{p}=(0,0,1)$ with the direction of the diboson system in the laboratory frame, $\hat{p}=\hat{p}_{ZZ}$, for the determination of the azimuthal decay angles. Both choices are equivalent in the $2\to 2$ picture as $\hat{p}_{ZZ}$ coincides with the beam direction. However, the directions are no longer equal in the presence of additional radiation, such as for NLO EW corrections. In~\autoref{fig:SMBPlots}, we compare the results obtained in these two frames and observe only minimal differences in the smallest eigenvalue.\footnote{In~\autoref{app:Basisdepend}, we present the effects due to this distinct frame choice to entanglement when considering the LO+PS and NLO QCD+PS samples. We observe that the alternative reference frame slightly degrades the entanglement marker for the considered observables.}

\section{$h\to 4\ell$}
\label{sec:hZZ}

 \begin{figure}[!h]
    \centering
    \includegraphics[width=0.8\textwidth]{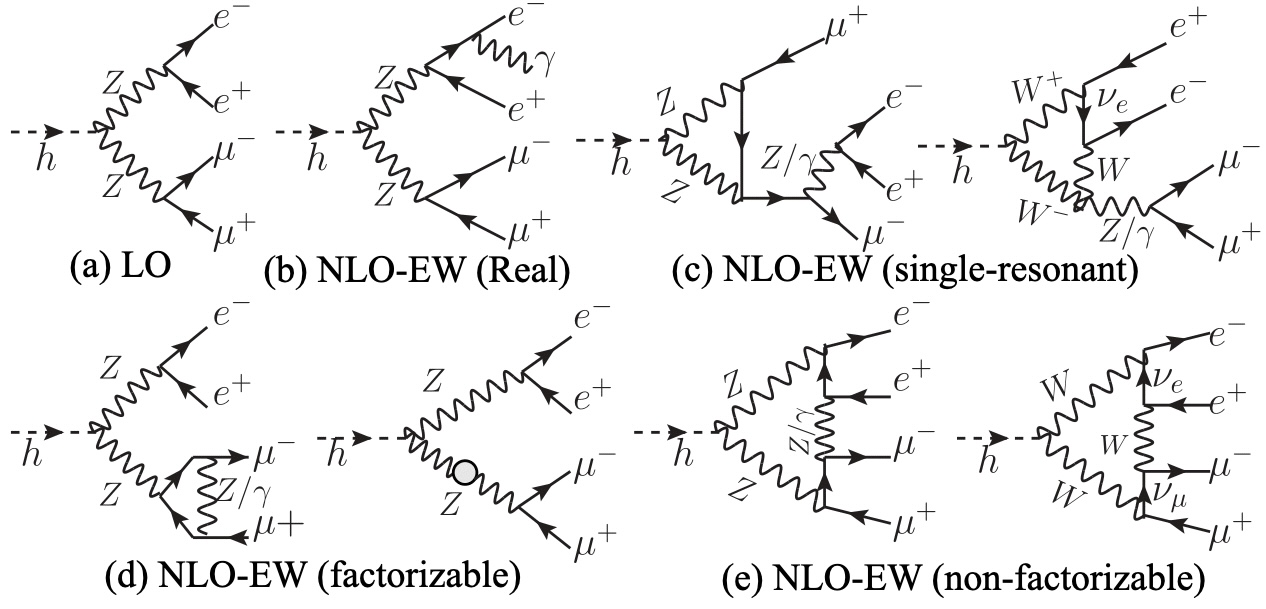}
    \caption{Representative sample of Feynman diagrams for the LO and NLO EW contributions to the Higgs boson decay into four charged leptons $h \rightarrow 4\ell$.}
    \label{fig:Feyn2}
\end{figure}

The SM Higgs boson decay $h \rightarrow e^+e^-\mu^+\mu^-$ is regarded as one of the most promising channels to detect quantum entanglement in a two-qutrit system at the LHC. See~\autoref{fig:Feyn2} for a representative sample of Feynman diagrams. Owing to the scalar nature of the Higgs boson, the $ZZ^\ast$ system has been shown in phenomenological studies to exhibit entanglement even when analyzed inclusively~\cite{Barr:2021zcp,Aguilar-Saavedra:2022wam,Fabbrichesi:2023cev,Fabbrichesi:2023jep,Fabbri:2023ncz,Bernal:2023ruk,Aguilar-Saavedra:2024whi,Subba:2024mnl,Bernal:2024xhm,Sullivan:2024wzl,Aguilar-Saavedra:2024jkj}.
Previous studies have considered this process at LO and demonstrated that, due to conservation of angular momentum and parity in the $h\to ZZ^\ast$ decay, the density matrix has a highly constrained structure, with only nine nonzero entries~\cite{Aguilar-Saavedra:2022wam}. In particular, at LO the following relations hold

\begin{align}
    A^{1}_{20}&=A^{2}_{20}\neq 0, \nonumber\\
    C_{1,1,1-1} &= C_{1,-1,1,1} = -C_{2,1,2,-1} = -C_{2,-1,2,1}\neq 0, \label{eq:LOConds}\\ 
    C_{2,2,2,-2}&= C_{2,-2,2,2} = -C_{1,0,1,0} = 2-C_{2,0,2,0}\neq 0, \nonumber\\
    \frac{A^{2,1}_{20}}{\sqrt{2}}+1 &= C_{2,2,2,-2}\nonumber,
\end{align}
and the density matrix has the form
\begin{align}
    \label{eq:rhoLOForm}
    \rho_{\mathrm{LO}} &= \begin{pmatrix}
        0 & 0 & 0 & 0 & 0 & 0 & 0 & 0 & 0 \\
        0 & 0 & 0 & 0 & 0 & 0 & 0 & 0 & 0 \\
        0 & 0 & \frac{1}{3}C_{2,2,2,-2} & 0 & \frac{1}{3}C_{2,1,2,-1} & 0 & \frac{1}{3}C_{2,2,2,-2} & 0 & 0 \\
        0 & 0 & 0 & 0 & 0 & 0 & 0 & 0 & 0 \\
        0 & 0 & \frac{1}{3}C_{2,1,2,-1} & 0 & \frac{1}{3}\left(3-2C_{2,2,2,-2}\right) & 0 & \frac{1}{3}C_{2,1,2,-1} & 0 & 0 \\
        0 & 0 & 0 & 0 & 0 & 0 & 0 & 0 & 0 \\
        0 & 0 & \frac{1}{3}C_{2,2,2,-2} & 0 & \frac{1}{3}C_{2,1,2,-1} & 0 & \frac{1}{3}C_{2,2,2,-2} & 0 & 0 \\
        0 & 0 & 0 & 0 & 0 & 0 & 0 & 0 & 0 \\
        0 & 0 & 0 & 0 & 0 & 0 & 0 & 0 & 0 \\
    \end{pmatrix}\,.
\end{align}
The coefficients $A_{LM}^i$ and $C_{L_1 M_1 L_2 M_2}$ are defined in \autoref{eq:Coeffs1} and \autoref{eq:Coeffs2}. Hence, at LO, nine nonzero elements in the density matrix depend on just two independent parameters, which can be represented as $C_{2,2,2,-2}$ and $C_{2,1,2,-1}$. 

While the Peres-Horodecki criterion typically provides only a sufficient condition for detecting entanglement in two-qutrit systems~\cite{PhysRevLett.77.1413,HORODECKI1997333}, the specific structure of the LO density matrix $\rho_\text{LO}$ in \autoref{eq:rhoLOForm} ensures that it also serves as a necessary condition~\cite{Aguilar-Saavedra:2022wam}.
Using~\autoref{eq:rhoLOForm}, the nonzero eigenvalues of $\rho^{T_2}$ are\footnote{The Peres-Horodecki criterion exploits the fact that, for a separable density matrix $\rho$, the partial transpose for a second subsystem
\begin{align}
    \rho^{T2}=\sum_i p_i\rho_A^i\otimes (\rho_B^i)^T\,,
\end{align}
yields in a non-negative operator. Thus, if $\rho^{T2}$ has at least one negative eigenvalue, the density matrix $\rho$ represents an entangled system.}
\begin{align}
    \pm\frac{1}{3}C_{2,1,2,-1},\,\pm\frac{1}{3}C_{2,2,2,-2},\,1-\frac{2}{3}C_{2,2,2,-2}.
\end{align}
Hence, these necessary and sufficient conditions for entanglement are satisfied when either $C_{2,2,2,-2}$ or $C_{2,1,2,-1}$ is nonzero~\cite{Aguilar-Saavedra:2022wam}. However, incorporating higher-order EW contributions may introduce novel contributions to the angular coefficients~\cite{DelGratta:2025qyp}, potentially disrupting these simple relations. It is therefore essential to investigate the impact of these corrections on the system. In~\autoref{fig:Feyn2}, we display a sample of representative Feynman diagrams for this process at LO and NLO EW. In this analysis, we focus on the on-shell Higgs boson decay. Given the Higgs boson mass of $m_h=125.09\, \mathrm{GeV}$~\cite{ATLAS:2023oaq}, one of the $Z$-bosons is necessarily off-shell. 

\subsection{Leading Order Decay}
\label{subsec:hZZ-LO}

\begin{figure}[!tb]
    \centering
    \includegraphics[width=0.55\textwidth]{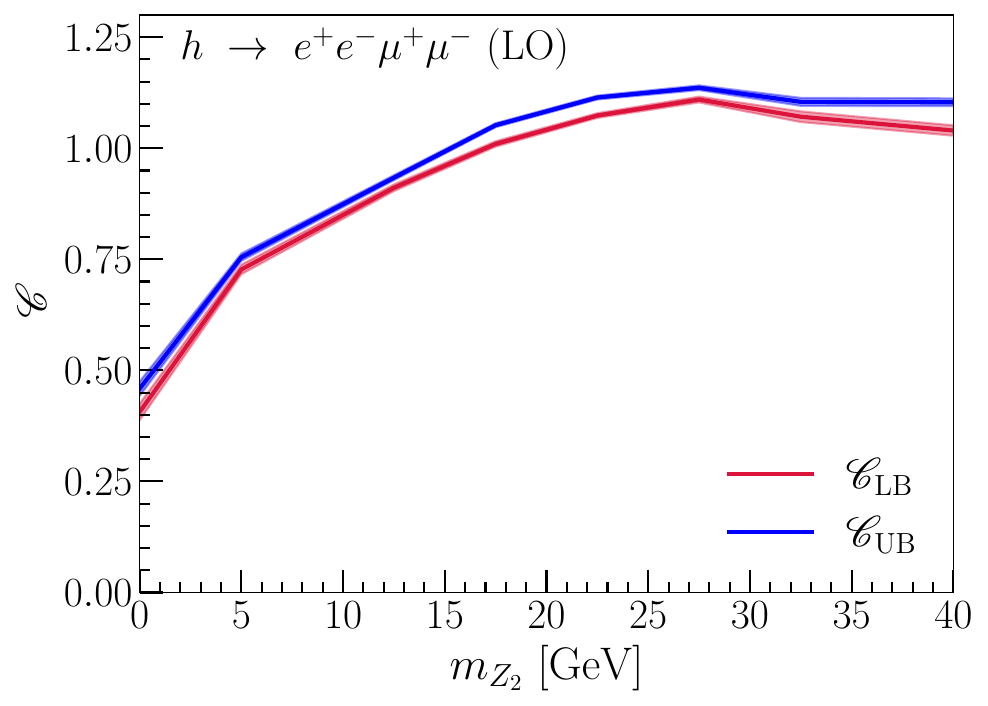}
    \caption{Lower $\mathscr{C}_{\mathrm{LB}}$ and upper $\mathscr{C}_{\mathrm{UB}}$ bounds of the concurrence for $h\to ZZ^\ast\to 4\ell$ (LO) as a function of the off-shell $Z$ mass.}
    \label{fig:h_ZZ_zmass}
\end{figure}

We begin our analysis with the LO results for this process. We generate events for $h\to ZZ^\ast\to e^+e^-\mu^+\mu^-$ with the \texttt{Prophecy4F} package~\cite{Denner:2019fcr}, which adopts the $G_F$ scheme. We use the same input parameters presented in~\cref{eq:mh,eq:mt,eq:CMparams1,eq:CMparams2,eq:GFparam}. Following the formalism outlined in~\autoref{sec:tomography}, we reconstruct the density matrix of the $ZZ^\ast$ system.
The values of the two independent coefficients at LO in the inclusive scenario are given by
\begin{align}
    C_{2,1,2,-1}&=-0.935(2)\,,\\
    C_{2,2,2,-2}&=0.579(1)\,.
\end{align}
The density matrix in this case is then given by
\begin{align}
\rho_{ZZ}^{\mathrm{LO}} &=
\begin{pmatrix}
 0 & 0 & 0 & 0 & 0 & 0 & 0 & 0 & 0 \\
 0 & 0 & 0 & 0 & 0 & 0 & 0 & 0 & 0 \\
 0 & 0 & 0.193(1) & 0 & -0.313(1) & 0 & 0.192(1) & 0 & 0 \\
 0 & 0 & 0 & 0 & 0 & 0 & 0 & 0 & 0 \\
 0 & 0 & -0.313(1) & 0 & 0.611(1) & 0 & -0.311(1) & 0 & 0 \\
 0 & 0 & 0 & 0 & 0 & 0 & 0 & 0 & 0 \\
 0 & 0 & 0.192(1) & 0 & -0.311(1) & 0 & 0.193(1) & 0 & 0 \\
 0 & 0 & 0 & 0 & 0 & 0 & 0 & 0 & 0 \\
 0 & 0 & 0 & 0 & 0 & 0 & 0 & 0 & 0 \\
\end{pmatrix}\,.
\label{eq:hZZdensincl}
\end{align}
This result agrees with the expected form from~\autoref{eq:rhoLOForm}, confirming that the relations from~\autoref{eq:LOConds} hold at LO. The zero entries indicate that the corresponding components are numerically compatible with zero within the uncertainties of the analysis.

In~\autoref{fig:h_ZZ_zmass}, we present the LO lower and upper bounds on the concurrence, $\mathscr{C}_{\mathrm{LB}}$ and $\mathscr{C}_{\mathrm{UB}}$, as a function of the lowest reconstructed mass for the same-flavor and opposite-sign dilepton, $m_{Z_2}$. The results show that both $\mathscr{C}_{\mathrm{UB}}$ and $\mathscr{C}_{\mathrm{LB}}$ remain positive across the entire distribution, signaling the presence of quantum entanglement. Furthermore, these concurrence bounds increase for larger $m_{Z_2}$. In this limit, the $ZZ^\ast$ approaches a pure state, showing maximum entanglement~\cite{Aguilar-Saavedra:2022wam,Fabbrichesi:2023cev}. 

\subsection{NLO EW Effects}
\label{subsec:hZZ-nloew}

\begin{figure}[t!]
    \centering
    \includegraphics[width=0.33\textwidth]{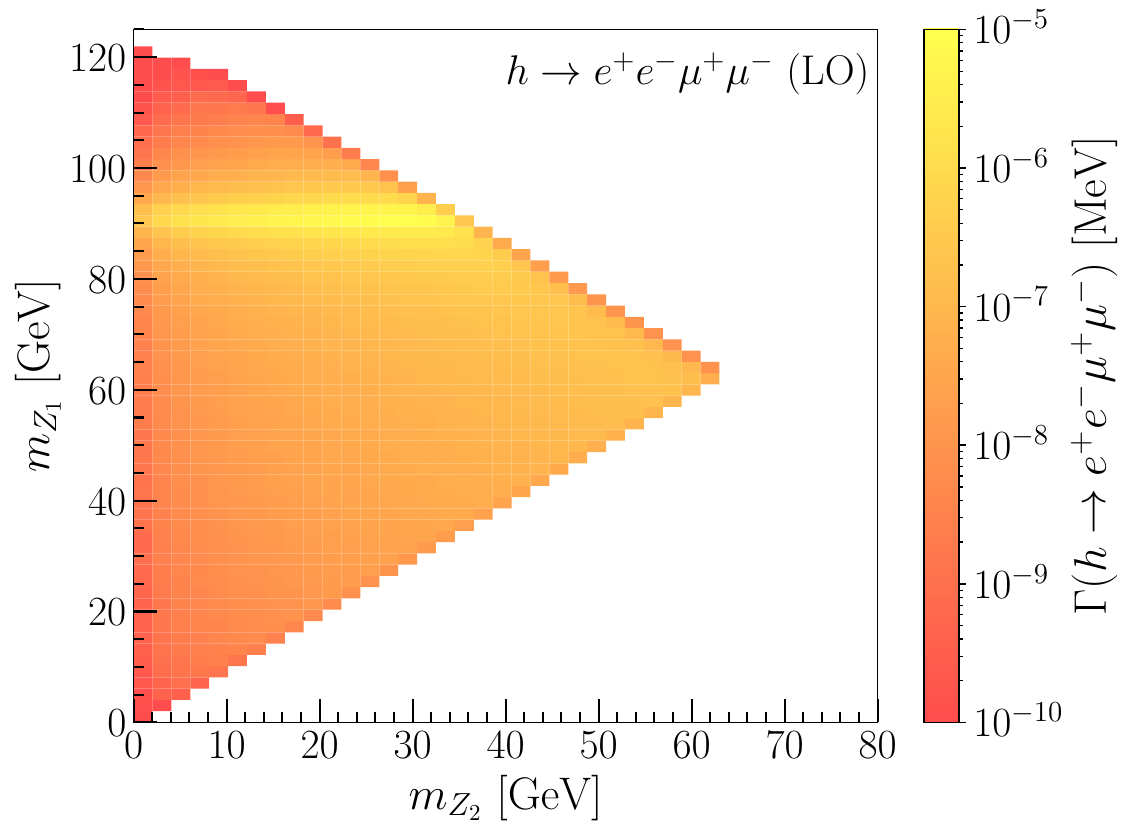}
    \includegraphics[width=0.33\textwidth]{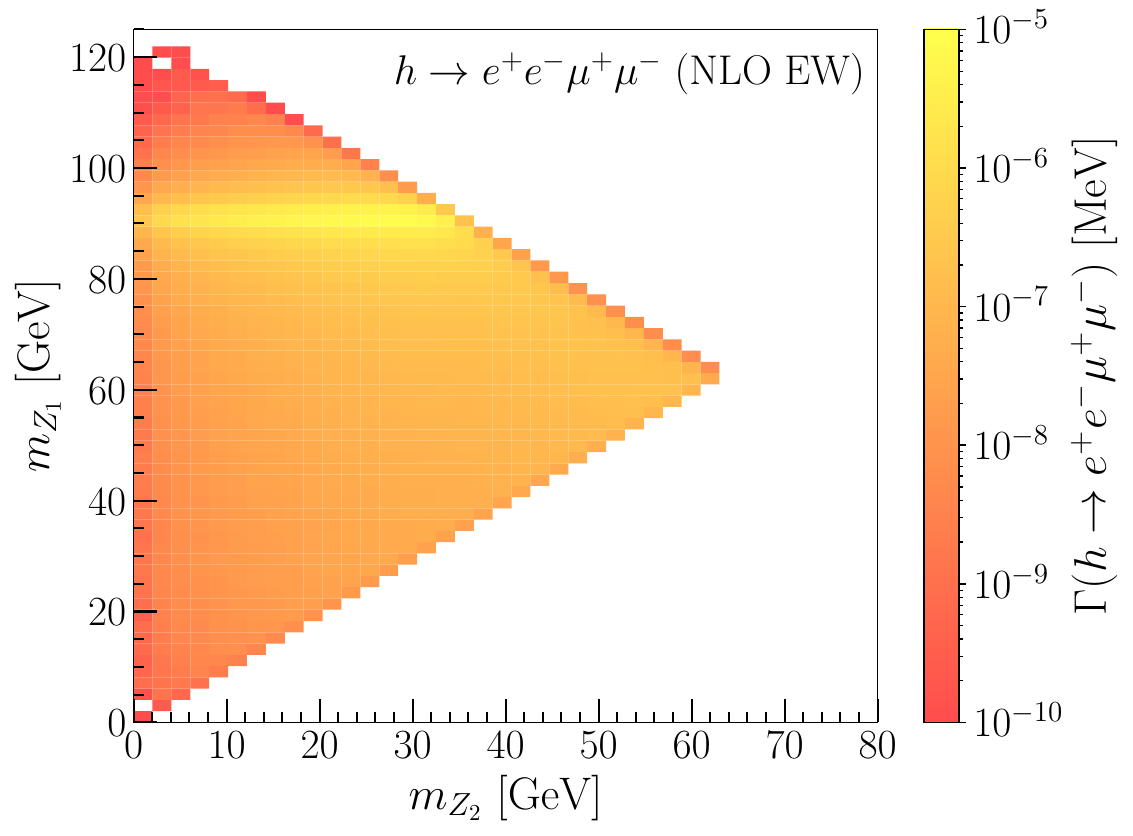}
    \includegraphics[width=0.31\textwidth]{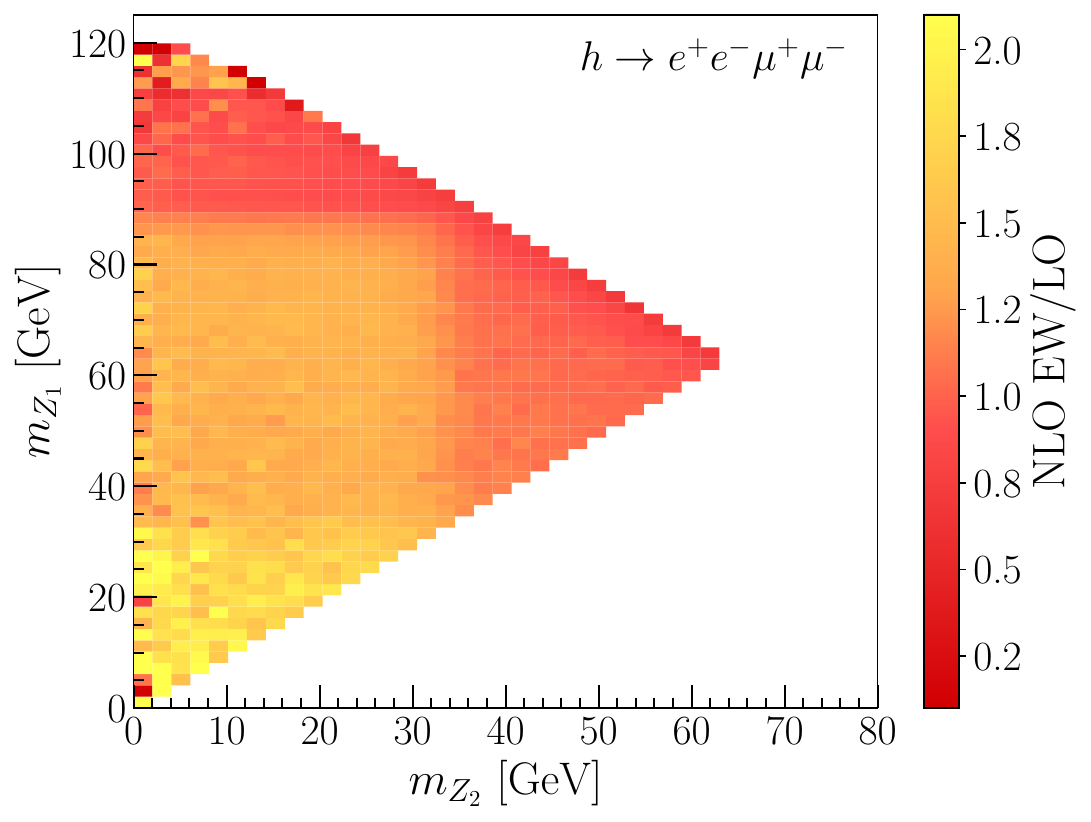}
    \caption{Decay distribution for $h \to e^+e^-\mu^+\mu^-$ at the LO (left panel), NLO EW (central panel), along with their corresponding ratios (right panel) in the $m_{Z_{1}}$-$m_{Z_{2}}$ plane.}
    \label{fig:h_ZZ_dist}
\end{figure}

To study the impact of the higher-order EW corrections, we generate events for $h\to e^+e^-\mu^+\mu^-$ at NLO EW with \texttt{Prophecy4F}~\cite{Denner:2019fcr}. The complex mass scheme is adopted~\cite{Denner:1999gp,Denner:2005fg} with the same input parameters and computational setup as in~\autoref{subsec:hZZ-LO}. Charged leptons and isolated photons are defined as in~\autoref{subsec:nlo-ew}. 
In~\autoref{fig:h_ZZ_dist}, we present the partial Higgs width for $h \to e^+e^-\mu^+\mu^-$ in the $m_{Z_{1}}$-$m_{Z_{2}}$ plane, where $m_{Z_{1}}$ and $m_{Z_{2}}$ represent the highest and lowest reconstructed same-flavor and opposite-sign dilepton masses, respectively. The left, central, and right panels correspond to the LO, NLO EW, and their ratio. The leading contributions to the Higgs partial width arise in the region with $80~\text{GeV}\lesssim m_{Z_1}\lesssim 100$~GeV and $10~\text{GeV}\lesssim m_{Z_2}\lesssim 40$~GeV. While the total NLO EW correction is relatively small, $\delta\Gamma(h\to e^+e^-\mu^+\mu^-)_\text{NLO}/\Gamma(h\to e^+e^-\mu^+\mu^-)_\text{LO}\sim 1.4\%$, it can become more pronounced in certain kinematic regions, leading to either enhanced or suppressed effects. As we demonstrate below, the structure of the NLO EW correction has important phenomenological implications for interpreting the data as a two-qutrit system.

At NLO EW, we find that the reconstructed density matrix using the procedure outlined in \autoref{sec:tomography} takes the following form:\footnote{The analysis is performed in the $4\ell$ rest frame using dressed leptons to define the helicity basis, as described in~\autoref{sec:tomography}.} 
\begin{align}
\resizebox{0.95\textwidth}{!}{$
\rho_{ZZ}^{\mathrm{NLO}}=\begin{pmatrix}
 0.102(4) & 0 & 0 & 0 & 0 & 0 & 0 & 0 & 0 \\
 0 & 0 & 0 & 0.103(4) & 0 & 0 & 0 & 0 & 0 \\
 0 & 0 & 0.100(4) & 0 & -0.207(4) & 0 & 0.188(1) & 0 & 0 \\
 0 & 0.103(4) & 0 & 0.003(1) & 0 & 0 & 0 & 0 & 0 \\
 0 & 0 & -0.207(4) & 0 & 0.591(1) & 0 & -0.199(4) & 0 & 0 \\
 0 & 0 & 0 & 0 & 0 & 0.004(1) & 0 & 0.109(4) & 0 \\
 0 & 0 & 0.188(1) & 0 & -0.199(4) & 0 & 0.100(4) & 0 & 0 \\
 0 & 0 & 0 & 0 & 0 & 0.109(4) & 0 & 0 & 0 \\
 0 & 0 & 0 & 0 & 0 & 0 & 0 & 0 & 0.098(4) \\
\end{pmatrix}\,.
$}
\label{eq:hZZdensNLO} 
\end{align}
Zero entries denote components that are compatible with zero within uncertainties, and only those differing from zero beyond the $1\sigma$ level are shown. In view of the sizable NLO EW effects to the electroweak mixing angle, as discussed in~\autoref{subsec:nlo-ew}, we use $\sin^2\theta_\text{eff}^\ell$ at one-loop to perform the quantum tomography, which leads to~\autoref{eq:hZZdensNLO} and underlies all subsequent results in this section, unless stated otherwise.\footnote{In \autoref{app:compare}, we present a comparison of our results with those of Ref.~\cite{DelGratta:2025qyp}, which investigates NLO EW corrections to Higgs decays and appeared during the final stages of our manuscript. We find numerical differences in some entries of the density matrix. These can be attributed to their use of $\sin\theta_W^{\mathrm{LO}}$ in the quantum tomography procedure for the NLO EW samples, rather than $\sin\theta^{\ell}_{\mathrm{eff}}$ as adopted in our work.} Compared to the LO density matrix, we observe that the NLO EW effects introduce nonzero values for the off-diagonal elements $\rho_{24}$ and $\rho_{68}$, as well as for the diagonal elements $\rho_{11}$, $\rho_{44}$, $\rho_{66}$, and $\rho_{99}$, as shown in~\autoref{eq:hZZdensNLO}. Several of these novel NLO elements are comparable in magnitude to the other nonzero elements of the LO density matrix. Additionally, the NLO EW corrections significantly affect the magnitude of the LO nonzero elements. 

\begin{table}[!t]
    \centering
    \begin{tabular}{|c|c|c|c|}
       \hline Coefficient & LO & \thead{NLO  \\ ($\sin^2\theta_{W}^{\mathrm{LO}}$)} & \thead{NLO  \\ ($\sin^2\theta_{\mathrm{eff}}^{\ell}$)} \\ \hline
        $A_{2,0}^{1}$ & $-0.5907(7)$ & $-0.5634(9)$ & $-0.5634(9)$ \\
        $A_{2,0}^{2}$ & $-0.5914(7)$ & $-0.5459(9)$  & $-0.5459(9)$ \\
        $C_{1,0,1,0}$ & $-0.580(8)$ & $0.00(1)$ & $0.00(1)$ \\
        $C_{1,1,1,-1}$ & $0.943(8)$ & $0.12(1)$ & $0.29(2)$ \\
        $C_{2,1,2,-1}$ &  $-0.935(2)$ & $-0.929(2)$ & $-0.929(2)$ \\
        $C_{2,0,2,0}$ & $1.413(2)$ & $1.374(2)$ & $1.374(2)$ \\
        $C_{2,2,2,-2}$ & $0.579(1)$ & $0.565(2)$ & $0.565(2)$ \\
        \hline 
    \end{tabular}
    \caption{The non-vanishing coefficients for the inclusive $h\to 4\ell$ decay at LO and NLO EW.  The LO coefficients are obtained applying quantum tomography with tree level weak mixing angle, while the NLO EW results are presented for two scenarios: one employing the LO mixing angle and the other using the effective one-loop mixing angle. The specific values of the mixing angles are listed in~\autoref{tab:sinw2}.}
    \label{tab:indcoeffsLONLO}
\end{table}

We find that when the reconstructed density matrix takes the form of~\autoref{eq:hZZdensNLO}, four of its nine eigenvalues are given by
\begin{align}
\frac{1}{2}\left(|\rho_{44}|\pm\sqrt{4|\rho_{24}|^2+|\rho_{44}|^2}\right)~\text{and}~\frac{1}{2}\left(|\rho_{66}|\pm\sqrt{4|\rho_{68}|^2+|\rho_{66}|^2}\right).
\label{eq:eigrhoNLO}
\end{align}
For $|\rho_{24}|\gg|\rho_{44}|$ and $|\rho_{68}|\gg|\rho_{66}|$, as numerically observed in~\autoref{eq:hZZdensNLO}, two of the eigenvalues of $\rho_\text{NLO}^{ZZ}$ are approximately given by $-|\rho_{24}|$ and $-|\rho_{68}|$. This result underscores that at NLO EW, the inclusive density matrix is guaranteed to have negative eigenvalues. Moreover, the magnitudes of these negative eigenvalues are directly set by $|\rho_{24}|$ and $|\rho_{68}|$. 
From Eqs.~(\ref{eq:hZZdensNLO}) and~(\ref{eq:eigrhoNLO}), we also observe that the magnitudes of these two eigenvalues are comparable. This similarity arises from the structure of the relevant matrix elements:
\begin{align}
    \rho_{24}=-\frac{1}{6}\left(C_{1,1,1,-1}+C_{1,1,2,-1}+C_{2,1,1,-1}+C_{2,1,2,-1} \right),\\ 
    \rho_{68}=-\frac{1}{6}\left(C_{1,1,1,-1}-C_{1,1,2,-1}-C_{2,1,1,-1}+C_{2,1,2,-1}\right),
\end{align}
where $C_{1,1,2,-1} = -0.01(1)$ and $C_{2,1,1,-1} = 0.02(1)$ are both consistent with zero within $2\sigma$, leading to $|\rho_{24}|\approx|\rho_{68}|\approx 0.103(4)$. Remarkably, these elements for the density matrix are identically zero at LO, since $C_{1,1,2,-1}=C_{2,1,1,-1}=0$ and $C_{1,1,1,-1}=-C_{2,1,2,-1}$, as shown in~\autoref{eq:LOConds}. Thus, the appearance of nonzero $\rho_{24}$ and $\rho_{68}$ signals that some of the LO relations from~\autoref{eq:LOConds} are not satisfied when including NLO EW corrections. In~\autoref{tab:indcoeffsLONLO}\footnote{There are sign differences in $C_{1,0,1,0}$ and $C_{1,1,1,-1}$ relative to Ref.~\cite{Grossi:2024jae} due to a different basis choice: whereas that work defines the $z$-axis separately for each $Z$, here it is defined with respect to $Z_1$ (see~\autoref{sec:tomography}), which leads to an overall sign flip for $\cos\theta_1$ and $\cos\theta_2$.}, we present the values of the non-vanishing coefficients at both LO and NLO EW in the inclusive scenario. Several of these coefficients receive sizable higher-order corrections, confirming the violation of the LO relations from~\autoref{eq:LOConds} at NLO EW.

It is important to understand to what extent each of the NLO EW contributions affects the quantum tomography approach presented in~\autoref{sec:tomography}. Virtual corrections, such as the single-resonant diagram in~\autoref{fig:Feyn2}~(c) and non-factorizable contribution in~\autoref{fig:Feyn2}~(e), indicate that the two-qutrit system may no longer be well-defined, invalidating the considered quantum tomography formalism as seen from~\autoref{eq:dsigma1}. Real corrections, as presented in~\autoref{fig:Feyn2}~(b), when the photon is not recombined, affect the form of the decay density matrix from~\autoref{eq:Gamma_mat}. Additionally, factorizable virtual corrections to the $Z\ell^{+}\ell^{-}$  vertex and self-energy diagrams, as illustrated in~\autoref{fig:Feyn2}~(d), modify the value of the weak mixing angle, which translates into a different spin analyzing power, $\eta_\ell$. Thus, applying the same quantum tomography procedure at LO and NLO EW may lead to unphysical results. 

As we discussed in~\autoref{subsec:nlo-ew}, the effects on the weak mixing angle stemming from contributions~\autoref{fig:Feyn2}~(d) can be effectively captured in the quantum tomography analysis, by replacing the tree level weak mixing angle with the one-loop effective weak mixing angle, $\sin^2\theta_\text{eff}^\ell$. This approach improves the description of the angular coefficients $A_{LM}^i$ and $C_{L_1,M_1,L_2,M_2}$ with $L=1$, whose determination relies on $\eta_\ell$ and are therefore sensitive to the weak mixing angle. The numerical impact of these corrections is visible in~\autoref{tab:indcoeffsLONLO}, where the coefficient $C_{1,1,1,-1}$ shows a sizable shift when extracted from NLO EW samples using $\sin^2\theta_W^\text{LO}$ versus $\sin^2\theta_\text{eff}^\ell$. 

\begin{figure}[!tb]
    \centering
    \includegraphics[width=0.33\textwidth]{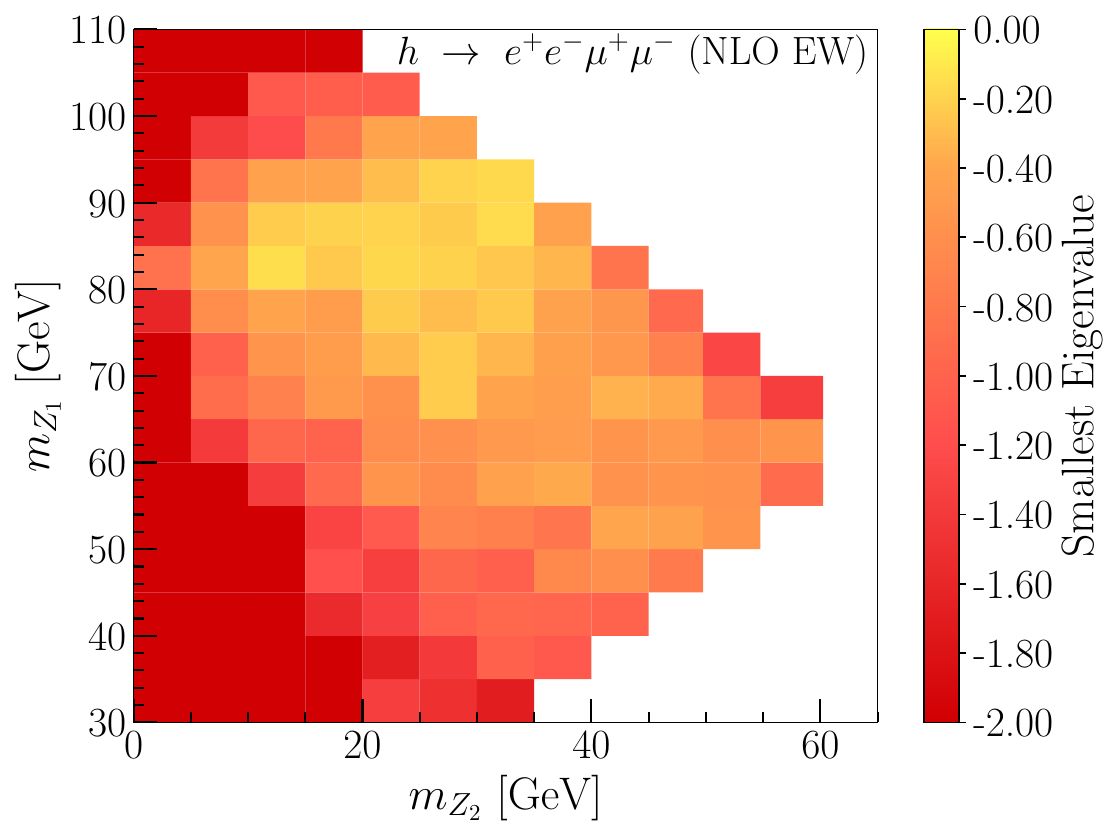}
    \includegraphics[width=0.33\textwidth]{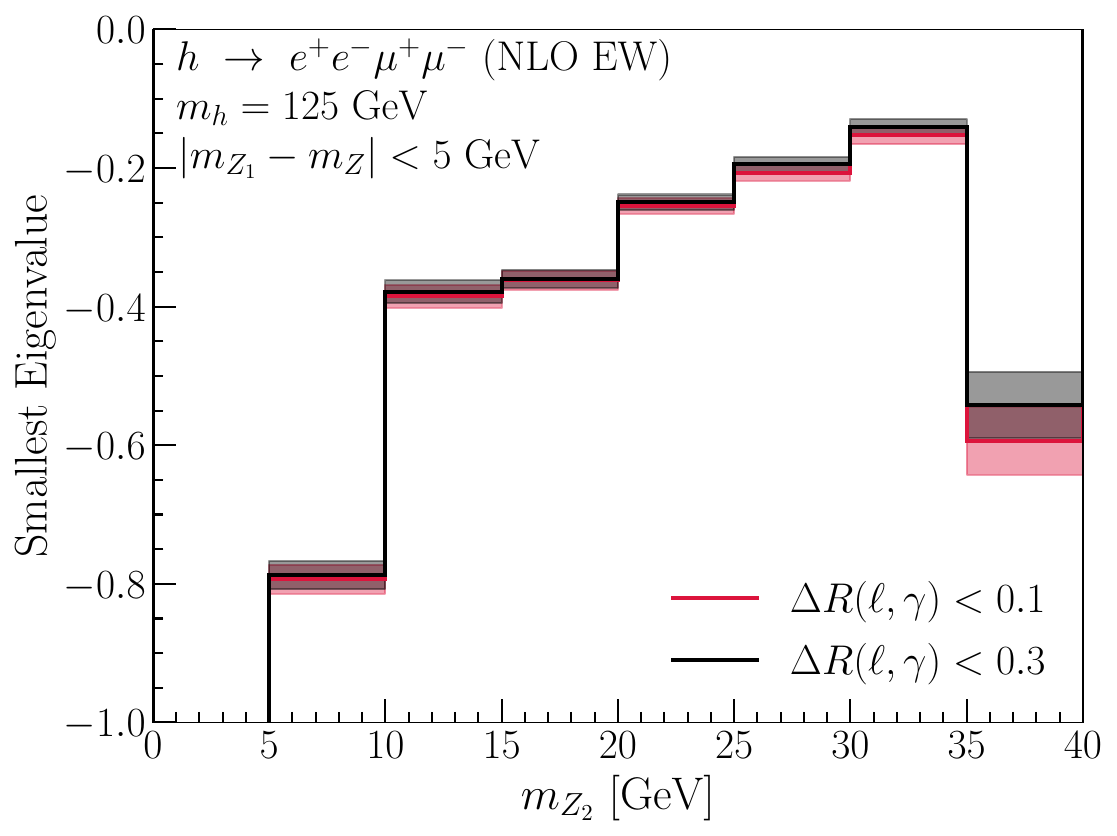}
    \includegraphics[width=0.33\textwidth]{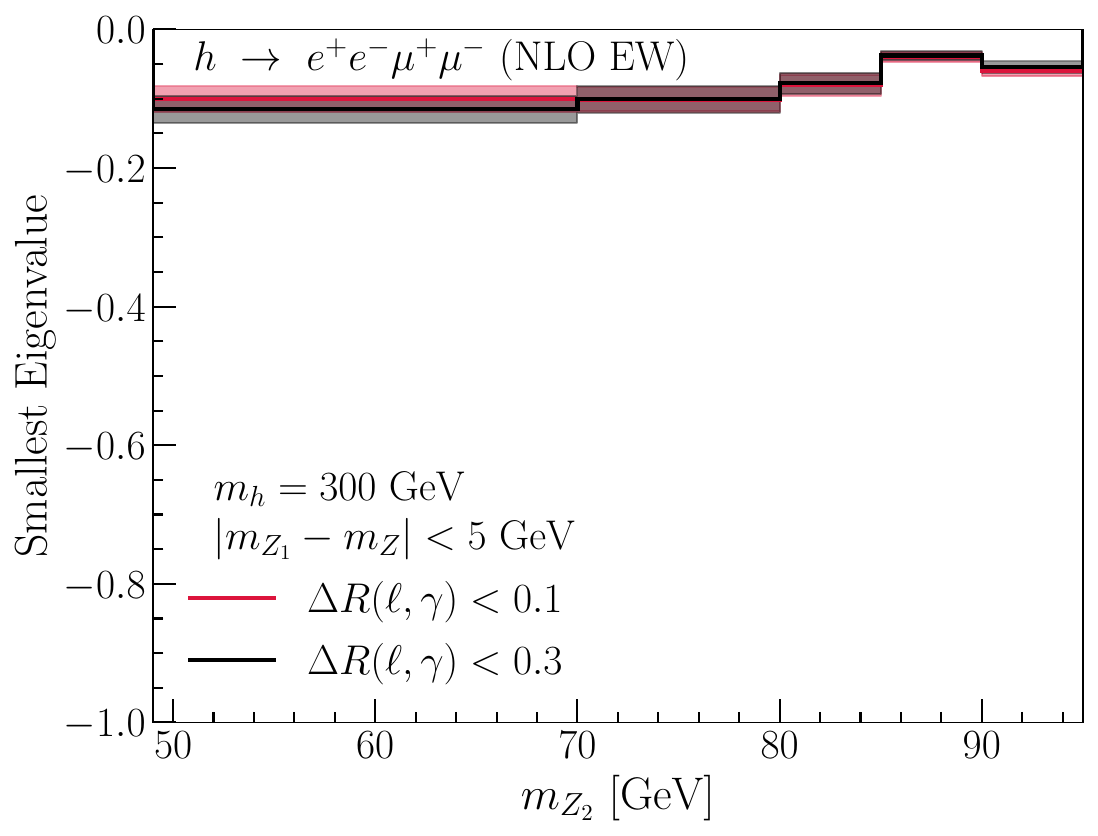}
    \caption{Smallest eigenvalue of the density matrix $\rho^{\mathrm{NLO}}_{ZZ}$ as a function of $(m_{Z_2},m_{Z_1})$ (left panel), and $m_{Z_2}$ for $m_h=125$~GeV (middle panel) and $m_h=300$~GeV (right panel). In the middle and right panels, we show the smallest eigenvalue for the recombination radii $\Delta R(\ell,\gamma)<0.1$ (red) and $\Delta R(\ell,\gamma)<0.3$ (black).} 
    \label{fig:h_ZZ_smallEig}
\end{figure}

In \autoref{fig:h_ZZ_smallEig} (left panel), we present the smallest eigenvalue of the reconstructed density matrix $\rho_\text{ZZ}^\text{NLO}$ as a function of $(m_{Z_2},m_{Z_1})$ invariant masses. We observe that the smallest eigenvalue is negative over the kinematically allowed phase space and becomes significantly more negative when the leading $Z_1$ boson is off-shell. In this kinematic regime, virtual corrections, such as those from non-factorizable corrections depicted in~\autoref{fig:Feyn2}~(e), can display relevant effects, undermining the interpretation of these results as a well-defined two-qutrit quantum system. While these contributions can be depleted enforcing an on-shell condition on the leading $Z$ boson, $|m_{Z_1} - m_Z| < 5$~GeV, which has a small impact on the partial decay width, the central panel of \autoref{fig:h_ZZ_smallEig} shows that this restriction is not sufficient to fully eliminate the negative eigenvalues. For more details on the angular coefficients in this kinematic regime, see~\autoref{tab:indcoeffsLONLO2}. 

\begin{table}[htbp]
    \centering
    \begin{tabular}{|c|c|c|c|}
       \hline Coefficient & LO & NLO & NLO/LO  \\ \hline
        $A_{2,0}^{1}$ & $-0.4910(8)$ & $-0.486(1)$ & $0.989$ \\
        $A_{2,0}^{2}$ & $-0.4915(8)$ & $-0.471(1)$ & $0.958$ \\
        $C_{1,0,1,0}$ & $-0.65(1)$ & $0.42(3)$ & $-0.646$ \\
        $C_{1,1,1,-1}$ & $1.008(9)$ & $-0.37(3)$ & $-0.279$ \\
        $C_{2,1,2,-1}$ &  $-0.993(2)$ & $-0.985(3)$ & $0.992$ \\
        $C_{2,0,2,0}$ & $1.335(2)$ & $1.328(4)$ & $0.995$ \\
        $C_{2,2,2,-2}$ & $0.646(2)$ & $0.635(3)$ & $0.983$ \\ \hline
    \end{tabular}
    \caption{The non-vanishing coefficients for the $h\to 4\ell$ decay at LO, NLO EW,  and their NLO/LO ratio. The invariant mass conditions $|m_{Z_1}-m_Z|<5$~GeV and $m_{Z_2}>10$~GeV are imposed. The LO coefficients are obtained with the tree level weak mixing angle and NLO EW with the effective weak mixing angle at one-loop.}
    \label{tab:indcoeffsLONLO2}
\end{table}

We highlight two key observations. First, enforcing the leading $Z$ boson to be on-shell does not restore the validity of the two-qutrit description. This breakdown stems from a combination of residual effects, including singly-resonant virtual corrections, such as those illustrated in~\autoref{fig:Feyn2}~(c), and real emission contributions, as shown in~\autoref{fig:Feyn2}~(b). Second, we observe a sign flip in two coefficients, $C_{1,1,1,-1}$ and $C_{1,0,1,0}$ from~\autoref{tab:indcoeffsLONLO2}, when comparing the LO and NLO EW results. This change of sign cannot be attributed to corrections to the spin analyzing power, since the extraction of both coefficients depends on $1/\eta_\ell^2$~\cite{Grossi:2024jae}.

Since $Z_1$ is produced near the $Z$ pole and $Z_2$ is typically off-shell with an invariant mass around $m_{Z_2} \sim 30$~GeV, one might consider whether the observed sign flip could be driven by the scale dependence of the weak mixing angle, with $\eta_\ell$ effectively absorbing these effects and allowing the NLO EW events to reproduce a density matrix similar to the LO scenario~\cite{DelGratta:2025qyp}. However, we find that the running of the effective weak mixing angle, $\sin^2\theta_\text{eff}^\ell(m_{Z_2})$, results in a sub-percent shift from its $Z$ pole value~\cite{SLACE158:2005uay,Erler:2004cx,Czarnecki:2000ic}, which translates into at most a $\delta\eta_\ell\approx 6$\% variation in the spin analyzing power for $Z_2$. Therefore, the approximation $\eta_{\ell}(m_{Z_1})\eta_{\ell}(m_{Z_2}) \approx \eta_\ell^2$, where $\eta_\ell \equiv \eta_\ell(m_Z)$, is well justified. The observed sign reversal in the $L = 1$ angular coefficients shown in~\autoref{tab:indcoeffsLONLO2} cannot be explained by the correction in $\eta_\ell$. Instead, it arises from other NLO effects, such as singly-resonant virtual corrections, which remain relevant because the 125~GeV Higgs mass ensures that at least one of the dilepton masses, $m_{Z_i}$, is off-shell.

 \begin{table}[!tbp]
    \centering
    \begin{tabular}{|c|c|c|c|}
       \hline Coefficient & LO & NLO  & NLO/LO\\ \hline
        $A_{2,0}^{1}$ & $-1.2131(8)$ & $-1.221(1)$ & $1.006$ \\
        $A_{2,0}^{2}$ & $-1.2120(8)$ & $-1.208(1)$ & $0.997$ \\
        $C_{1,0,1,0}$ & $-0.151(8)$ & $-0.06(2)$ & $0.397$ \\
        $C_{1,1,1,-1}$ & $0.623(9)$ & $0.45(3)$ & $0.722$ \\
        $C_{2,1,2,-1}$ &  $-0.612(1)$ & $-0.610(2)$ & $0.997$ \\
        $C_{2,0,2,0}$ & $1.843(2)$ & $1.849(3)$ & $1.003$ \\
        $C_{2,2,2,-2}$ & $0.132(2)$ & $0.145(3)$ & $1.098$ \\ \hline
    \end{tabular}
    \caption{The non-vanishing coefficients for the $h\to 4\ell$ decay for $m_h=300$~GeV at LO, NLO EW, and their NLO/LO ratio. The invariant mass conditions $|m_{\ell^{+}\ell^{-}}-m_Z|<5$~GeV is imposed for both $Z_1$ and $Z_2$.}
    \label{tab:indcoeffsmh300}
\end{table}

In the following, we study in more detail the effect of real emission and singly-resonant virtual corrections. To assess the relevance of the real emission, we compute the smallest eigenvalue of the reconstructed density matrix, as shown in the central panel of \autoref{fig:h_ZZ_smallEig}, while dressing the charged leptons with photons using isolation cones of varying radii, $\Delta R(\ell, \gamma) < 0.1$ and $0.3$. We find that the smallest eigenvalue does not change significantly\footnote{We note that the smallest eigenvalue remains significantly negative, indicating that the density matrix is not positive semi-definite at NLO EW. Hence, the two-qutrit interpretation breaks down, and concurrence bounds or other entanglement indicators, as computed at LO, are no longer well-defined. For a comparison of the NLO EW corrections in the $h\to e^+e^-\mu^+\mu^-$ and $h \to WW^\ast \to \ell^+ \nu_\ell \ell'^- \bar{\nu}_{\ell'}$ systems, see Ref.~\cite{Goncalves:2025xer}.}. To assess the effects of singly-resonant virtual corrections on the Higgs decay, we generate new LO and NLO EW samples, imposing a larger Higgs mass of $m_h=300$~GeV. At this higher mass, both $Z_1$ and $Z_2$ can be close to the $Z$ pole, effectively suppressing the non-resonant and singly-resonant virtual corrections. Since the effective weak mixing angle, $\sin^2\theta_\text{eff}^\ell$, receives one-loop corrections that depend on the Higgs mass, we update its value accordingly using the \texttt{GRIFFIN} package~\cite{Chen:2022dow} to carry out the quantum tomography study.\footnote{The effective weak mixing angle at NLO for a Higgs mass of $m_h = 300~\rm{GeV}$ is $\sin^2 \theta_{\rm eff}^\ell = 0.231659$. The Higgs-mediated correction contributes only a sub-dominant shift to the weak mixing angle~\cite{Sirlin:2012mh,Awramik:2006uz,Chiesa:2019nqb},  with $\frac{\sin^2 \theta_{\rm eff}^\ell(m_h = 300~\rm{GeV})}{\sin^2 \theta_{\rm eff}^\ell(m_h = 125~\rm{GeV})} = 0.997$.} In~\autoref{fig:h_ZZ_smallEig} (right panel), we present the smallest eigenvalue for the reconstructed density matrix as a function of $m_{Z_2}$, imposing the on-shell condition $|m_{Z_1} - m_Z| < 5$~GeV. We find that the deviation of the smallest eigenvalue from zero is significantly reduced compared to the SM Higgs scenario with $m_h=125$~GeV, particularly when both $Z_1$ and $Z_2$ are close to the $Z$ pole, which dominates the considered partial Higgs decay width for $m_h = 300$~GeV. Remarkably, in the $m_h=300$~GeV  scenario, higher-order corrections to the angular coefficients are strongly suppressed, as shown in \autoref{tab:indcoeffsmh300}.\footnote{Although the double-resonant contribution dominates for $m_h=300\, \rm{GeV}$, some non-factorizable effects remain and can be further suppressed by the $5\,\rm{GeV}$ selection around $m_Z$, see~\autoref{fig:h_ZZ_smallEig} (right panel).} In particular, the sign flip previously observed for $C_{1,1,1,-1}$ and $C_{1,0,1,0}$ in the $m_h = 125$~GeV case no longer appears, and the results more closely resemble those of the LO prediction. This supports the interpretation that the sign reversal in the SM Higgs scenario originates from singly-resonant contributions. Thus, the Higgs decay $h\to e^+e^-\mu^+\mu^-$ in the SM receives sizable NLO EW corrections to the angular coefficients, particularly from singly-resonant contributions, as the double-resonant $ZZ$ topology is kinematically forbidden. As discussed in this section, the structure and magnitude of these higher-order effects challenge the interpretation of this Higgs decay as a two-qutrit $ZZ$ in the SM framework.

\section{Summary}
\label{sec:summary}

Polarization and spin correlations in diboson systems offer sensitive probes for both precision studies and searches for new physics. Recently, the interpretation of these angular observables through the lens of quantum information, such as assessing whether the diboson system is entangled, has brought a compelling new perspective to these studies~\cite{Barr:2021zcp,Aguilar-Saavedra:2022wam,Aoude:2023hxv,Fabbrichesi:2023cev,Fabbrichesi:2023jep,Fabbri:2023ncz,Bernal:2023ruk,Morales:2023gow,Bi:2023uop,Barr:2024djo,Aguilar-Saavedra:2024whi,Subba:2024mnl,Bernal:2024xhm,Sullivan:2024wzl,Aguilar-Saavedra:2024jkj,Grossi:2024jae,DelGratta:2025qyp}. 

In this work, we performed a detailed study of the angular coefficients for $pp \to e^+e^-\mu^+\mu^-$ and $h\to e^+e^-\mu^+\mu^-$,  incorporating off-shell effects as well as higher-order QCD and electroweak corrections. A precise understanding of these effects is crucial for informing precision measurements and new physics studies. Guided by the fundamental properties of the spin density matrix (hermiticity, unit trace, and positive semi-definiteness), we investigate whether the two-qutrit system remains physically well-defined in the presence of higher-order corrections. While the tree level structure of gauge boson decays is fully captured by a spherical harmonic expansion up to rank-2, realistic analysis requires accounting for off-shell and higher-order corrections, possibly introducing higher-rank terms. These effects can undermine the completeness of the expansion and challenge the validity of the two-qutrit interpretation.

For $pp \to e^+e^-\mu^+\mu^-$, we find that NLO QCD corrections significantly affect the angular observables, as novel production channels, such as $qg$ initial states, modify the helicity structure of the process. While these corrections preserve a well-defined two-qutrit description, they substantially weaken the strength of the entanglement indicators. However, by binning events in jet multiplicity, we find that the exclusive zero-jet sample enhances the entanglement marker and aligns more closely with the LO expectation. Unlike QCD effects, the NLO EW corrections impact both production and decay and give rise to non-factorizable contributions. Whereas restricting the invariant mass window around the $Z$ pole mitigates some of these effects, angular coefficients with $L=1$ still receive sizable corrections.  They can be traced to stem primarily from radiative corrections to the weak mixing angle. Thus, a more accurate quantum tomography for this process requires the inclusion of higher-order corrections to the weak mixing angle, which can be systematically incorporated by replacing the tree level value with the one-loop effective weak mixing angle, $\sin^2\theta_\text{eff}^\ell$.

For the Higgs boson decay to four leptons, the LO analysis yields a tightly constrained structure for the angular coefficients, featuring only a few nonzero angular coefficients and constraining relations among them. At tree level, the system displays a well-defined density matrix for a two-qutrit system, which shows entanglement even for inclusive event samples. However, NLO EW corrections, especially those from singly-resonant contributions, introduce large corrections that disrupt this LO picture. Remarkably, the magnitude and structure of these higher-order effects challenge the interpretation of this Higgs decay as a two-qutrit system in the SM framework. This tension is evidenced by the presence of sizable negative eigenvalues for the reconstructed density matrix. 

In summary, our results underscore the importance of careful treatment of higher-order (i.e., quantum) effects when studying quantum observables at colliders, such as quantum entanglement. Sizable radiative corrections can substantially modify entanglement indicators and, in some cases, undermine the interpretation of systems that may appear well-defined at leading order. Therefore, incorporating these corrections yields more reliable theoretical predictions and enables more precise and informed experimental analyses.

\section*{Acknowledgments}
We thank Werner Bernreuther for clarifications regarding the basis definition in Ref.~\cite{Bernreuther:2015yna}, Emanuele Re for helping us with the \texttt{POWHEG-BOX-RES}, and Ayres Freitas for valuable discussions on higher-order corrections to the weak mixing angle in the \texttt{GRIFFIN} package~\cite{Chen:2022dow}. The work of DG, AK, and AN is supported in part by US Department of Energy Grant Number DE-SC 0016013. Some computing for this project was performed at the High Performance Computing Center at Oklahoma State University, supported in part through the National Science Foundation grant OAC-1531128. 
F.K. gratefully acknowledges support by STFC under grant agreement ST/P006744/1.
\appendix

\section{Irreducible Tensor Operators}
\label{app:TensOp}

The density matrix for a system of two vector bosons, as introduced in~\autoref{eq:rho}, is parametrized in terms of irreducible tensor operators $T^{L_1}_{M_1},\,T^{L_2}_{M_2} \in \left\{\mathbb{I}_3,~T_{-1}^1,~T_0^1,~T_1^1,~T_{-2}^2,~T_{-1}^{2},~T_0^2,~T_1^2,~T_2^2 \right\}\,$~\cite{Aguilar-Saavedra:2022wam}. These operators can be expressed in terms of the spin-1 component operators, $J_x$, $J_y$, and $J_z$ as

\begin{align}
    T_{\pm1}^1 &= \mp\frac{\sqrt{3}}{2}\left(J_x\pm i~J_y\right)\,, && T_0^1 = \sqrt{\frac{3}{2}}J_z, \nonumber\\
    T_{\pm 2}^2 &= \frac{2}{\sqrt{3}}\left( T_{\pm 1}^1\right)^2\,, && T_{\pm 1}^2 = \sqrt{\frac{3}{2}}\left( T_{\pm 1}^1 T_0^1 + T_0^1 T_{\pm 1}^1\right)\,, \\
    T_{0}^2 &= \frac{\sqrt{2}}{3}\left( T_1^1 T_{-1}^1 + T_{-1}^1 T_1^1 + 2(T_0^1)^2\right)\nonumber\,.
\end{align}

Explicitly,
\begin{align}
    T_{-1}^1 = \sqrt{\frac{3}{2}}\begin{pmatrix}
        0 & 0 & 0 \\
        1 & 0 & 0 \\
        0 & 1 & 0
    \end{pmatrix}\,, && T_0^1 = \sqrt{\frac{3}{2}}\begin{pmatrix}
        1 & 0 & 0 \\
        0 & 0 & 0 \\
        0 & 0 & -1
    \end{pmatrix}\,, && T_1^1 = \sqrt{\frac{3}{2}}\begin{pmatrix}
        0 & -1 & 0 \\
        0 & 0 & -1 \\
        0 & 0 & 0
    \end{pmatrix}\,,
\end{align}

and

\begin{align}
    T_{-2}^2 &= \sqrt{3}\begin{pmatrix}
        0 & 0 & 0 \\
        0 & 0 & 0 \\
        1 & 0 & 0 
    \end{pmatrix}\,, && T_{-1}^2 = \sqrt{\frac 3 2}\begin{pmatrix}
        0 & 0 & 0 \\
        1 & 0 & 0 \\
        0 & -1 & 0 
    \end{pmatrix}\,, && T_{0}^2 = \frac{1}{\sqrt{2}}\begin{pmatrix}
        1 & 0 & 0 \\
        0 & -2 & 0 \\
        0 & 0 & 1 
    \end{pmatrix}\,,\\
    T_{1}^2 &= \sqrt{\frac 3 2}\begin{pmatrix}
        0 & -1 & 0 \\
        0 & 0 & 1 \\
        0 & 0 & 0 
    \end{pmatrix}\,, && T_{2}^2 = \sqrt{3}\begin{pmatrix}
        0 & 0 & 1 \\
        0 & 0 & 0 \\
        0 & 0 & 0 
    \end{pmatrix}\,.
\end{align}

\section{NLO Fixed-Order vs. NLO+PS Matched Predictions}
\label{app:compPS}

\begin{table}[th!]
\resizebox{\textwidth}{!}{
    \centering
    \begin{tabular}{|c|c|c|c|c |}
       \hline Coefficient & Fixed-order NLO QCD & NLO QCD+PS & Fixed-order NLO EW & NLO EW+PS\\ \hline
        $A_{1,0}^{1}$ & $-0.03(1)$ & $-0.035(3)$ &$0.004(3)$ & $0.004(3)$\\
        $A_{2,0}^{1}$ & $0.313(4)$ & $0.301(1)$  &$0.326(1)$ & $0.325(1)$\\
        $A_{2,1}^{1}$ & $0.109(3)$ & $0.1023(7)$ &$0.0949(7)$ & $0.0949(6)$\\
        $A_{1,0}^{2}$ & $0.026(8)$ & $0.024(2)$  &$-0.002(3)$ & $-0.003(3)$\\
        $A_{2,0}^{2}$ & $0.307(7)$ & $0.299(1)$  &$0.338(1)$ & $0.338(1)$ \\
        $A_{2,1}^{2}$ & $0.121(3)$ & $0.1165(7)$ &$0.1017(7)$ &$0.1018(7)$ \\
        $C_{1,0,1,0}$ & $0.7(1)$ & $0.75(1)$     &$0.63(3)$ & $0.62(3)$ \\
        $C_{2,0,2,0}$ & $0.195(8)$ & $0.198(3)$  &$0.234(3)$ & $0.235(3)$ \\
        $C_{2,2,2,2}$ & $-0.337(5)$ & $-0.332(2)$ &$-0.381(2)$ & $-0.381(2)$\\ \hline
    \end{tabular}
    }
    \caption{Comparison of the angular coefficients for $pp \to e^{+}e^{-}\mu^{+}\mu^{-}$ at fixed-order NLO QCD and fixed-order NLO EW with those obtained from NLO QCD+PS and NLO EW matched to QCD+QED parton shower. The selection $|m_{\ell^{+}\ell^{-}} - m_Z| < 10$ GeV is imposed to isolate the on-shell $Z$ contribution. Statistical uncertainties are quoted in parentheses. The imaginary parts of the coefficients are consistent with zero within uncertainties.}
    \label{tab:coeffsFOPS}
\end{table}

In this appendix, we compare fixed-order predictions with those obtained after parton shower matching, considering the $pp\to 4\ell$ channel. This complements the NLO QCD results in~\autoref{subsec:nlo-qcd} and the NLO EW results in~\autoref{subsec:nlo-ew}, which were presented including parton shower effects. 
In~\autoref{tab:coeffsFOPS}, we show a subset of angular coefficients evaluated at fixed-order NLO QCD and fixed-order NLO EW, and compare them respectively with NLO QCD+PS and NLO EW with QCD+QED PS (as in~\autoref{subsec:nlo-qcd} and~\autoref{subsec:nlo-ew}). The results show good agreement in both cases, confirming that the differences between LO and NLO results highlighted in the main text arise from genuine fixed-order corrections rather than from parton shower effects. The mild impact of the shower is expected, as it primarily modifies the kinematics, whereas the fixed-order corrections drive the effects on the helicity amplitudes of $ZZ$ production.

\section{Frame Dependence in $pp \to 4\ell$}
\label{app:Basisdepend}

\begin{figure}[!tbh]
    \centering
    \includegraphics[width=0.45\textwidth]{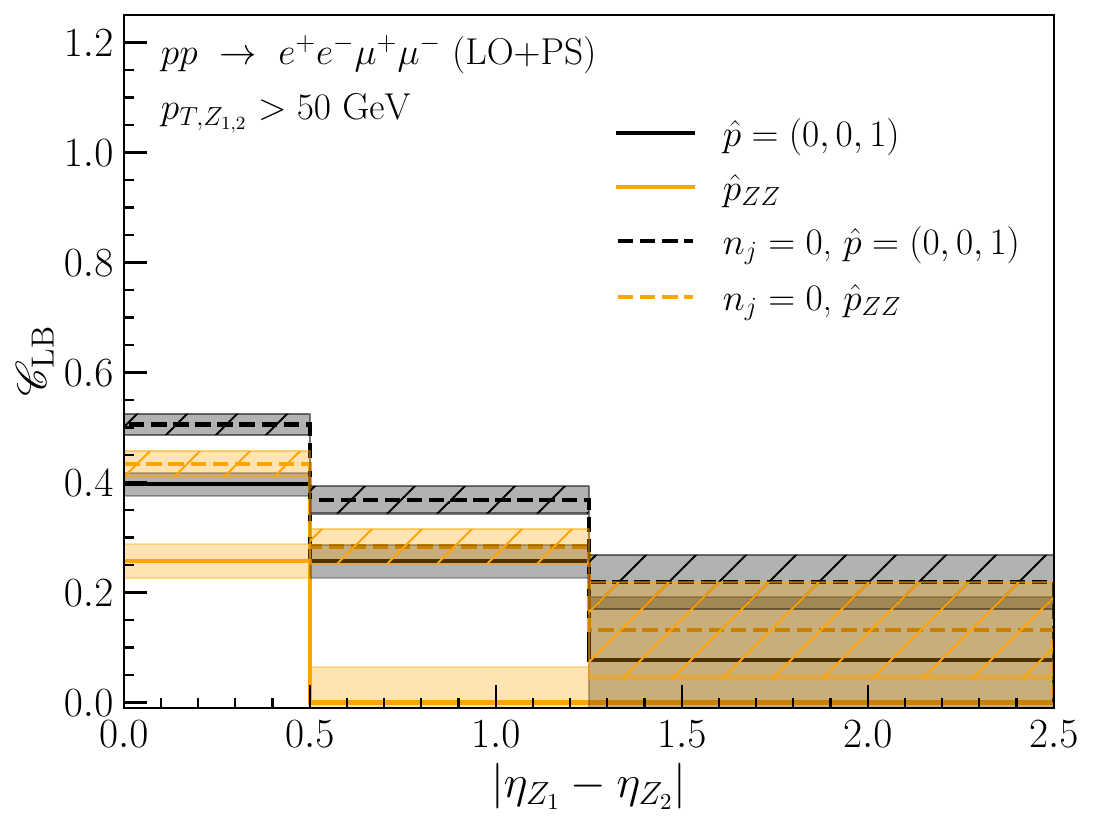}
    \includegraphics[width=0.45\textwidth]{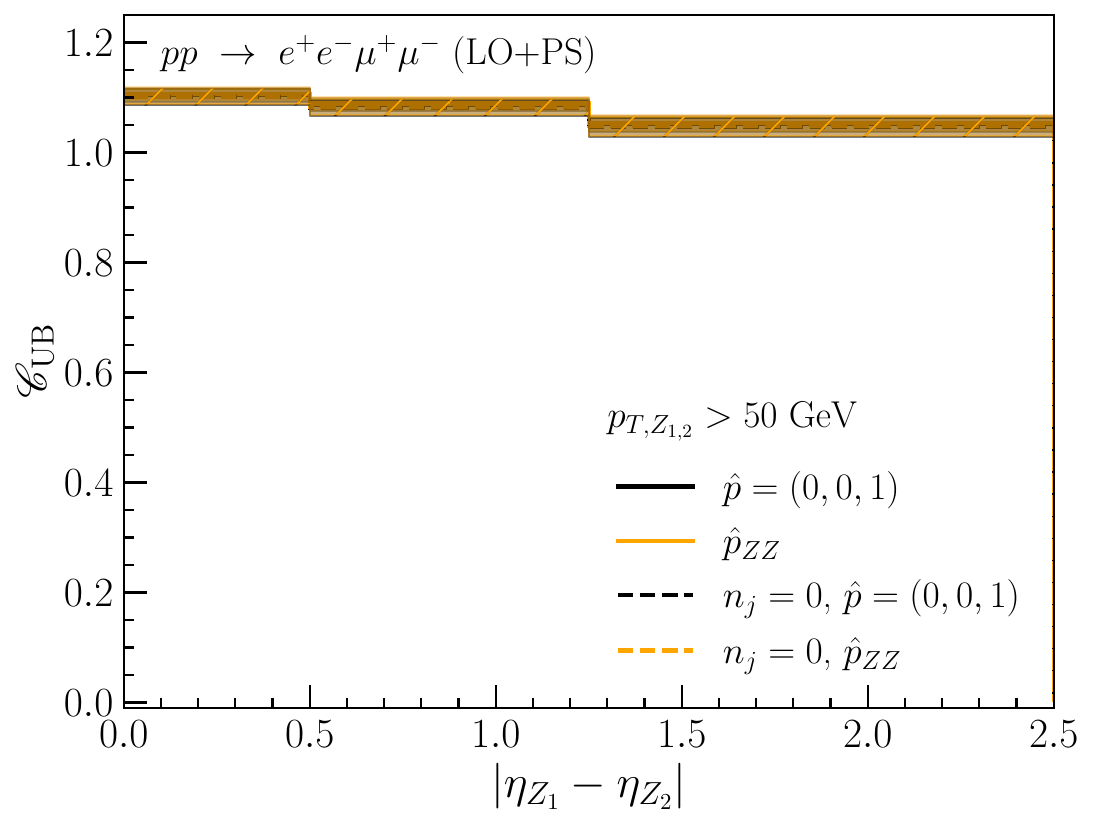} \\
    \includegraphics[width=0.45\textwidth]{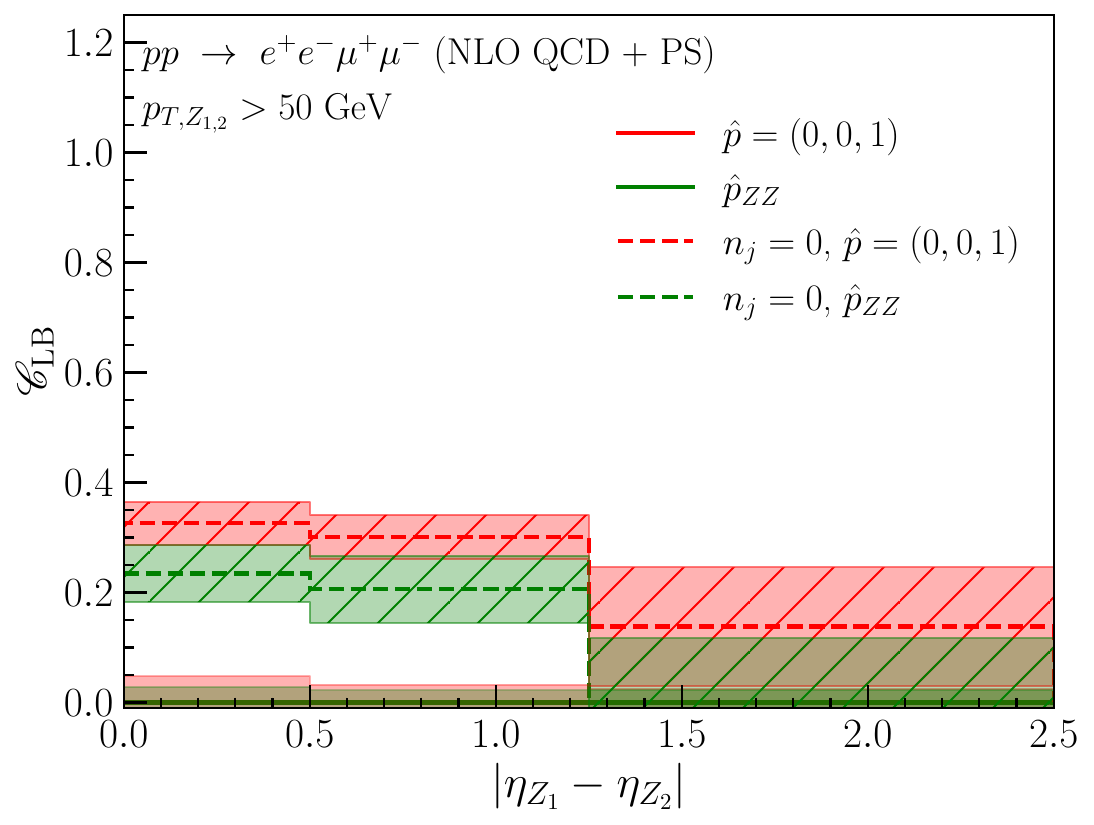}
    \includegraphics[width=0.45\textwidth]{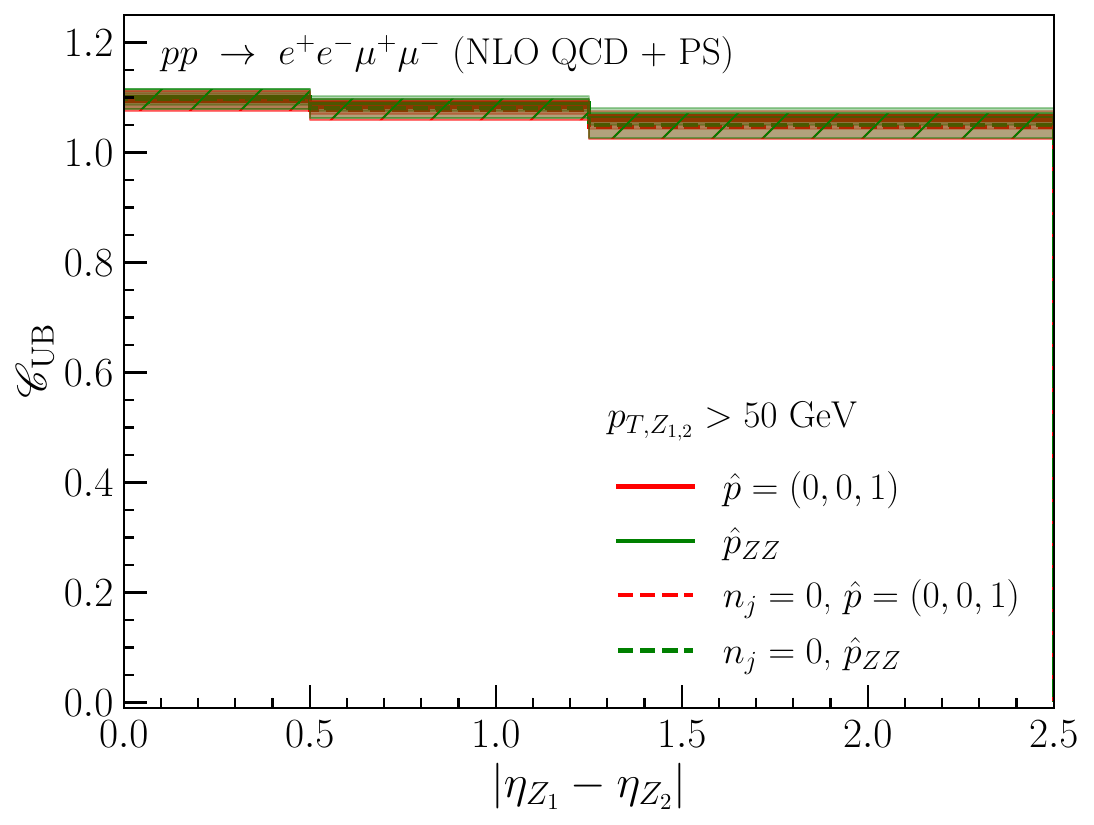}
    \caption{Comparison of concurrence bounds $\mathscr{C}_{\rm{LB}}$ (left panel) and $\mathscr{C}_{\rm{UB}}$ (right panel) in two frames with different reference axis $\hat{p}=(0,0,1)$ and $\hat{p}=\hat{p}_{ZZ}$, for $pp\to e^{+}e^{-}\mu^{+}\mu^{-}$ at LO+PS (upper panel) and NLO QCD+PS (lower panel). The concurrence bounds are shown as a function of $|\eta_{Z_1}-\eta_{Z_2}|$ with $p_{T,Z_{(1,2)}}>50\,\rm{GeV}$. The jet-inclusive samples are shown in solid lines, while the $n_j=0$ are shown with dashed lines. The error band accounts for both statistical and theoretical errors. The theoretical error is estimated as the maximum deviation from the central value $\mu_R=\mu_F=m_Z$ when varying the renormalization and factorization scales to $m_Z/2$ and $2m_Z$.}
    \label{fig:CboundsNLO_Compa}
\end{figure}

Since the angular coefficients depend on the reference frame used to reconstruct the lepton angles, we examine how the entanglement measure depends on the choice of reference frame. In particular, we compute the bounds on concurrence in the helicity frame, see~\autoref{sec:tomography}, with those obtained in the alternative frame employed in Ref.~\cite{Grossi:2024jae}. This alternative frame replaces the reference axis $\hat{p}=(0,0,1)$ with the direction of the diboson system in the laboratory frame, $\hat{p}=\hat{p}_{ZZ}$, for the determination of the azimuthal decay angles. Although both definitions are equivalent in a $2\to 2$ scattering, as the $ZZ$ direction coincides with the beam axis, they differ in the presence of additional radiation. In~\autoref{fig:CboundsNLO_Compa}, we show the lower (left panel) and upper (right panel) bounds of the concurrence, $\mathscr{C}_{\rm{LB}}$  and $\mathscr{C}_{\rm{UB}}$, for $pp\to e^{+}e^{-}\mu^{+}\mu^{-}$ at LO+PS (upper panel) and NLO QCD+PS (lower panel) as a function of $|\eta_{Z_1}-\eta_{Z_2}|$ with  the transverse momentum cuts
$p_{T,Z_{(1,2)}}>50\,\rm{GeV}$.  We observe that $\mathscr{C}_{\rm{LB}}$ is degraded in the $\hat{p}_{ZZ}$ basis for the inclusive jet samples, showing better agreement for the $n_j=0$ samples. $\mathscr{C}_{\rm{UB}}$ remains approximately stable in both frames.

\section{Comparison for $pp\to 4\ell$ and $h\to 4\ell$}
\label{app:compare}

The impact of NLO QCD and EW effects on the angular coefficients for $ZZ$ production was also studied in Ref.~\cite{Grossi:2024jae}. To validate the work presented in the main text, we have generated events for $pp\to 4\ell$ at LO, NLO QCD, and NLO EW using the same input parameters and basis used in Ref.~\cite{Grossi:2024jae}.\footnote{For a detailed comparison of how the choice of frame affects the entanglement measure in $pp\to 4\ell$ channel, between our frame and the alternative frame employed in Ref.~\cite{Grossi:2024jae}, see~\autoref{app:Basisdepend}.}  The comparison is presented in~\autoref{tab:coeffComp}, showing agreement within uncertainties.

\begin{table}[h]
\resizebox{\textwidth}{!}{
    \centering
    \begin{tabular}{|c|ccc|ccc|}\hline
       \multirow{2}{*}{Coefficient}  & \multicolumn{3}{c|}{This work} & \multicolumn{3}{c|}{Ref.~\cite{Grossi:2024jae}} \\ 
       \cline{2-7}
         & LO & NLO QCD & NLO EW & LO & NLO QCD & NLO EW \\\hline\hline
        $\alpha_{1,0}^{(1)}$ & $-0.0002(3)$ & $-0.0012(3)$ & $ -0.00013(9)$ & $-0.00001(9)$ & $-0.00097(10)$ & $-0.00004(9)$ \\ 
        $\alpha_{2,0}^{(1)}$ & $0.0293(3)$ & $0.0274(3)$ &$0.0284(9)$ & $0.03009(11)$ & $0.02794(13)$ & $0.02960(11)$ \\
        $\alpha_{1,0}^{(3)}$ & $-0.0001(3)$ & $-0.0009(3)$ & $0.00008(9)$ & $0.00012(13)$ & $-0.00086(15)$ & $0.00018(14)$ \\ 
        $\alpha_{2,0}^{(3)}$ & $0.0293(3)$ & $0.0272(3)$ & $0.0295(9)$& $0.03006(7)$ & $0.02796(6)$ & $0.02964(7)$ \\
        $\gamma_{1,0,1,0}$ & $-0.0018(1)$ & $-0.0014(1)$ & $-0.0005(3)$ & $-0.00173(3)$ & $-0.00148(3)$ & $-0.00043(3)$ \\
        $\gamma_{2,0,2,0}$ & $0.0019(1)$ & $0.00163(9)$ & $0.0018(3)$& $0.00188(2)$ & $ 0.00168(2)$ & $0.00187(2)$ \\
        $\alpha_{2,-2}^{(1)}$ & $-0.0095(2)$ & $-0.0099(2)$ & $ -0.0110(6)$& $-0.00967(7)$ & $-0.00993(9)$ & $-0.00991(7)$ \\
        $\alpha_{2,-2}^{(3)}$ & $-0.0096(2)$ & $-0.0095(2)$ & $-0.0112(6)$ & $-0.00973(4)$ & $-0.01003(4)$ & $-0.00996(4)$ \\ \hline
    \end{tabular}
    }
    \caption{Comparison of the angular coefficients for $pp\to 4\ell$ at LO, NLO QCD, and NLO EW between this work and Ref.~\cite{Grossi:2024jae} for the inclusive scenario. To perform this validation, events were generated using the same input parameters and basis as in Ref.~\cite{Grossi:2024jae}. The $\alpha$ and $\gamma$ coefficients are related to our $A$ and $C$ coefficients through: $\sqrt{40\pi}\alpha^{(i)}_{2,M} = A^{(i)}_{2,M}$, $40\pi\gamma_{2,M_1,2,M_2} = C_{2,M_1,2,M_2}$, and $8\pi\gamma_{1,M_1,1,M_2}/\eta_\ell^2 = C_{1,M_1,1,M_2}$}
    \label{tab:coeffComp}
\end{table}

While finalizing this work, Ref.~\cite{DelGratta:2025qyp} presented a related study on the impact of the higher-order corrections to $h \to 4\ell$. To enable a meaningful comparison with our computational setup, we generated new event samples for $h \to 4\ell$ at both LO and NLO EW using \texttt{Prophecy4F}~\cite{Denner:2019fcr}, adopting the same input parameters as those used in Ref.~\cite{DelGratta:2025qyp}. We implemented the quantum tomography procedure using the LO weak mixing angle, $\sin\theta_W^{\mathrm{LO}}$, even for the NLO EW sample, to match their analysis. The resulting LO inclusive density matrices from Ref.~\cite{DelGratta:2025qyp} and from our numerical analysis are shown in \autoref{eq:densLO1} and \autoref{eq:densLO2}, respectively. We find good agreement between the two calculations, with differences within the quoted statistical uncertainties (shown in parentheses).

\begin{align}
\rho^{\mathrm{LO}} &=
\begin{pmatrix}
 0 & 0 & 0 & 0 & 0 & 0 & 0 & 0 & 0 \\
 0 & 0 & 0 & 0 & 0 & 0 & 0 & 0 & 0 \\
 0 & 0 & 0.195(2) & 0 & -0.313(3) & 0 & 0.194(1) & 0 & 0 \\
 0 & 0 & 0 & 0 & 0 & 0 & 0 & 0 & 0 \\
 0 & 0 & -0.313(3) & 0 & 0.612(1) & 0 & -0.313(3) & 0 & 0 \\
 0 & 0 & 0 & 0 & 0 & 0 & 0 & 0 & 0 \\
 0 & 0 & 0.194(1) & 0 & -0.313(3) & 0 & 0.195(3) & 0 & 0 \\
 0 & 0 & 0 & 0 & 0 & 0 & 0 & 0 & 0 \\
 0 & 0 & 0 & 0 & 0 & 0 & 0 & 0 & 0 \\
\end{pmatrix}\,,
\label{eq:densLO1}
\end{align}

\begin{align}
\rho^{\mathrm{LO}} &=
\begin{pmatrix}
 0 & 0 & 0 & 0 & 0 & 0 & 0 & 0 & 0 \\
 0 & 0 & 0 & 0 & 0 & 0 & 0 & 0 & 0 \\
 0 & 0 & 0.194(3) & 0 & -0.305(4) & 0 & 0.193(1) & 0 & 0 \\
 0 & 0 & 0 & 0 & 0 & 0 & 0 & 0 & 0 \\
 0 & 0 & -0.305(4) & 0 & 0.610(1) & 0 & -0.313(3) & 0 & 0 \\
 0 & 0 & 0 & 0 & 0 & 0 & 0 & 0 & 0 \\
 0 & 0 & 0.193(1) & 0 & -0.313(3) & 0 & 0.193(3) & 0 & 0 \\
 0 & 0 & 0 & 0 & 0 & 0 & 0 & 0 & 0 \\
 0 & 0 & 0 & 0 & 0 & 0 & 0 & 0 & 0 \\
\end{pmatrix}\,.
\label{eq:densLO2}
\end{align}

The NLO EW inclusive density matrices from Ref.~\cite{DelGratta:2025qyp} and from this work are shown in~\autoref{eq:densNLO1} and~\autoref{eq:densNLOeff}, respectively. Unlike the LO case, we observe discrepancies for some entries. These differences can be traced to the use of $\sin\theta_W^{\mathrm{LO}}$ in the definition of $\eta_\ell$ in Ref.~\cite{DelGratta:2025qyp}, instead of the effective weak mixing angle $\sin\theta^{\ell}_{\mathrm{eff}}$ adopted in this work. This is confirmed by the density matrix shown in~\autoref{eq:densNLO2}, where we perform the quantum tomography using the LO weak mixing angle for the NLO EW sample. The resulting $\rho^{\mathrm{NLO}}_{\sin\theta_W^{\mathrm{LO}}}$ agrees with~\autoref{eq:densNLO1} within uncertainties. We note, however, that factorizable virtual corrections to the $Z\ell^+\ell^-$ vertex and self-energies, illustrated in~\autoref{fig:Feyn2}~(d), shift the weak mixing angle and consequently the spin analyzing power $\eta_\ell$. Therefore, applying the same quantum tomography setup at LO and NLO EW may lead to unphysical results. Further details are provided in~\autoref{subsec:nlo-ew} and~\autoref{subsec:hZZ-nloew}.

\begin{align}
\resizebox{\textwidth}{!}{$
\rho^{\mathrm{NLO}}=\begin{pmatrix}
 0.099(4) & 0 & 0 & 0 & 0 & 0 & 0 & 0 & 0 \\
 0 & 0.004(2) & 0 & 0.131(4) & 0 & 0 & 0 & 0 & 0 \\
 0 & 0 & 0.111(4) & 0 & -0.183(4) & 0 & 0.189(1) & 0 & 0 \\
 0 & 0.131(4) & 0 & -0.009(2) & 0 & 0 & 0 & 0 & 0 \\
 0 & 0 & -0.183(4) & 0 & 0.591(1) & 0 & -0.183(4) & 0 & 0 \\
 0 & 0 & 0 & 0 & 0 & -0.009(2) & 0 & 0.131(4) & 0 \\
 0 & 0 & 0.189(1) & 0 & -0.183(4) & 0 & 0.110(3) & 0 & 0 \\
 0 & 0 & 0 & 0 & 0 & 0.131(4) & 0 & 0.004(2) & 0 \\
 0 & 0 & 0 & 0 & 0 & 0 & 0 & 0 & 0.099(3) \\
\end{pmatrix}\,.
$}
\label{eq:densNLO1}
\end{align}

\begin{align}
\resizebox{\textwidth}{!}{$
\rho^{\mathrm{NLO}}_{\sin\theta^{\ell}_{\mathrm{eff}}}=\begin{pmatrix}
 0.09(1) & 0 & 0 & 0 & 0 & 0 & 0 & 0 & 0 \\
 0 & 0.001(5) & 0 & 0.08(1) & 0 & 0 & 0 & 0 & 0 \\
 0 & 0 & 0.11(1) & 0 & -0.23(1) & 0 & 0.189(3) & 0 & 0 \\
 0 & 0.08(1) & 0 & -0.006(5) & 0 & 0 & 0 & 0 & 0 \\
 0 & 0 & -0.23(1) & 0 & 0.596(2) & 0 & -0.23(1) & 0 & 0 \\
 0 & 0 & 0 & 0 & 0 & -0.003(4) & 0 & 0.08(1) & 0 \\
 0 & 0 & 0.189(2) & 0 & -0.23(1) & 0 & 0.11(1) & 0 & 0 \\
 0 & 0 & 0 & 0 & 0 & 0.08(1) & 0 & -0.002(5) & 0 \\
 0 & 0 & 0 & 0 & 0 & 0 & 0 & 0 & 0.09(1) \\
\end{pmatrix}\,.
$}
\label{eq:densNLOeff}
\end{align}

\begin{align}
\resizebox{\textwidth}{!}{$
\rho^{\mathrm{NLO}}_{\sin\theta_W^{\mathrm{LO}}}=\begin{pmatrix}
 0.097(6) & 0 & 0 & 0 & 0 & 0 & 0 & 0 & 0 \\
 0 & 0.001(5) & 0 & 0.127(6) & 0 & 0 & 0 & 0 & 0 \\
 0 & 0 & 0.106(6) & 0 & -0.184(6) & 0 & 0.189(3) & 0 & 0 \\
 0 & 0.127(6) & 0 & -0.002(4) & 0 & 0 & 0 & 0 & 0 \\
 0 & 0 & -0.184(6) & 0 & 0.594(2) & 0 & -0.185(6) & 0 & 0 \\
 0 & 0 & 0 & 0 & 0 & -0.003(4) & 0 & 0.131(4) & 0 \\
 0 & 0 & 0.189(3) & 0 & -0.185(6) & 0 & 0.102(7) & 0 & 0 \\
 0 & 0 & 0 & 0 & 0 & 0.125(7) & 0 & 0.007(3) & 0 \\
 0 & 0 & 0 & 0 & 0 & 0 & 0 & 0 & 0.096(7) \\
\end{pmatrix}\,.
$}
\label{eq:densNLO2}
\end{align}

\end{sloppypar}
\bibliographystyle{utphys}

\bibliography{reference}
\end{document}